\documentclass[preprint]{aastex701}
\usepackage{booktabs}
\usepackage{amsmath}
\usepackage{longtable}

\usepackage{hyperref}

\begin{document}

\title{Rebrightening in the Radio: Secondary Flares from Tidal Disruption Events AT2019ehz and AT2019eve}

\author[orcid=0009-0005-0038-4141]{N. K. Velez}
\affiliation{Department of Physics and Astronomy, University of Oregon, 1371 E 13th Ave, Eugene OR 97403, USA}
\email{nvelez@uoregon.edu}  

\author[orcid=0000-0001-7007-6295]{Y. Cendes}
\affiliation{Department of Physics and Astronomy, University of Oregon, 1371 E 13th Ave, Eugene OR 97403, USA}
\email{yncendes@uoregon.edu}  

\author[orcid=0009-0000-9210-8964]{J. Lynch}
\affiliation{Department of Physics and Astronomy, University of Oregon, 1371 E 13th Ave, Eugene OR 97403, USA}
\email{jlyn@uoregon.edu}  

\author[orcid=0000-0002-8297-2473]{K. D. Alexander}
\affiliation{Department of Astronomy and Steward Observatory, University of Arizona, 933 North Cherry Avenue, Tucson, AZ 85721-0065, USA}
\email{kdalexander@arizona.edu}

\author[orcid=0000-0002-9392-9681]{E. Berger}
\affiliation{Center for Astrophysics | Harvard \& Smithsonian, Cambridge, MA 02138, USA}
\email{eberger@cfa.harvard.edu}

\author[orcid=0000-0002-7706-5668]{R. Chornock}
\affiliation{Department of Astronomy, University of California, Berkeley, CA 94720-3411, USA}
\affiliation{Berkeley Center for Multi-messenger Research on Astrophysical Transients and Outreach (Multi-RAPTOR), University of California, Berkeley, CA 94720-3411, USA}
\email{chornock@berkeley.edu}

\author[orcid=0000-0003-3441-8299]{A. J. Goodwin}
\affiliation{International Centre for Radio Astronomy Research – Curtin University, GPO Box U1987, Perth, WA 6845, Australia}
\email{ajgoodwin.astro@gmail.com}

\author[orcid=0000-0003-1792-2338]{T. Laskar}
\affiliation{Department of Physics \& Astronomy, University of Utah, Salt Lake City, UT 84112, USA}
\email{tanmoy.laskar@utah.edu}

\author[orcid=0000-0003-4768-7586]{R. Margutti}
\affiliation{Department of Astronomy, University of California, Berkeley, CA 94720-3411, USA}
\affiliation{Berkeley Center for Multi-messenger Research on Astrophysical Transients and Outreach (Multi-RAPTOR), University of California,
Berkeley, CA 94720-3411, USA}
\affiliation{Department of Physics, University of California, 366 Physics North MC 7300, Berkeley, CA 94720, USA}
\email{rmargutti@berkeley.edu}

\begin{abstract}

We present the discovery of new secondary delayed radio flares from two tidal disruption events (TDEs), AT2019ehz and AT2019eve from observations ranging 1--18 GHz using the Karl G. Jansky Very Large Array.  For AT2019ehz, we find a secondary flare in radio emission began at 1545 days post-disruption, 770 days after the first radio flare detected at 775 days post-disruption, and for AT2019eve the secondary flare was detected at 1594 days post-disruption, 825 days after its initial radio detection at 769 days post-disruption.  Through spectral modeling of the multi-frequency observations, we extract physical parameters of the new flares to constrain the origin of the rebrightening. We find the secondary flare of AT2019ehz is likely produced by a second outflow launched at roughly 1000 days.  For AT2019eve, the secondary flare reveals unexpected spectral evolution with a self-absorbed peak increasing from 1325 to 2173 days post-disruption, which is not well explained with most models of expanding outflows and instead suggests a more complex physical geometry.  This inconclusive origin of AT2019eve highlights the need for more complex models to explain observed TDE radio evolution.

\end{abstract}


\section{Introduction} \label{sect:intro}
\setcounter{footnote}{0}

Tidal disruption events (TDEs) offer rare and valuable laboratories for investigating accretion processes, jet formation, and revealing and characterizing supermassive black holes (SMBHs) that were previously quiescent. TDEs occur when a star passes inside the tidal radius of a SMBH where the tidal forces then exceed the star’s own self-gravity \citep{hills75-power}. As a result roughly half of the star’s mass becomes unbound and escapes the system, while the rest of the debris settles into bound orbits that circularize around the SMBH and ultimately form an accretion disk \citep{rees88-td}. 

TDEs are most commonly discovered by their flares in optical/UV/ or X-ray (e.g. \citealt{vanVelzen21-tdediscs}) with ${\sim}$200 TDEs known to date \citep[see ][]{otter}. After discovery, follow-up radio monitoring of these TDEs reveal 20-30\% of the population exhibit prompt flares of radio emission from synchrotron radiation created by subsequent non-collimated outflows or jets interacting with the surrounding environment \citep{alexander20-radprops}. In recent years, it was discovered that approximately 40\% of TDEs show an onset of this radio emission at late-times ($> $2 years post-discovery; \citealt{cendes24-mega, horesh21-delayed}). Since then, systematic monitoring of TDEs at these late times continues to reveal new behaviors that remain poorly understood, such as the observation of multiple subsequent radio flares (e.g. ASASSN-15oi, \citealt{hajela25-15oi, horesh21-delayed}; AT2020vwl, \citealt{goodwin25-20vwl2};  AT2020zso, \citealt{christy2025-20zso/21sdu}; AT2024tvd, \citealt{sfaradi25-24tvd}).  This new flare of emission, or ``rebrightening," occurs on varying timescales in comparison to the initial flares. For example, in the case of ASSASN-15oi, radio emission was first detected at 182 days post-disruption and rose again at 1741 days with a rise in flux at a rate of ${\sim} t^4$ from the previous observation \citep{horesh21-delayed,hajela25-15oi}. AT2020vwl was first observed to display radio emission at 142 days with a rebrightening detected between 400-600 days post-disruption, rising at $F_{\nu}{\sim} t^{3.5}-t^{7.5}$ across all frequencies \citep{goodwin23-20vwl,goodwin25-20vwl2}. Another notable example is AT2024tvd, which is the fastest evolving TDE in this sample, displaying initial radio emission at 131 days with a sharp rise of in flux of $ t^{18}$ at 194 days \citep{sfaradi25-24tvd}.
The origin of TDE rebrightenings is poorly understood.  Potential origins of these flares include interactions of the original outflow with a denser circumnuclear medium, a re-injection of energy into the existing outflow, a secondary outflow being launched at later times, or a repeated partial TDE \citep{matsumoto24-bondi,zhuang25-clouds,matsumoto23-offaxis, zhong22-pTDErates}.

Here we present the continued monitoring of TDEs AT2019ehz and AT2019eve, which we find are exhibiting secondary outflow behavior. Previously reported activity for these TDEs are detailed in Section \ref{sect:past}. In Section \ref{sect:observations}, we report observations and data reduction procedures. The sources' light curves are analyzed in Section \ref{sect:light curve}. We present the results of our model fitted to the spectral data in Section \ref{sect:spectral fitting} and discuss the outflow modeling in Section \ref{sect:outflow modeling}. In Section \ref{sect:discussion}, we discuss the possible origins of the emission. We summarize our findings in Section \ref{sect:conclusion}.

\section{Backgrounds and Previous Observations} \label{sect:past}
AT2019ehz was optically discovered on 2019 April 29 at $z$ = 0.074 \citep{vanVelzen21-tdediscs}. X-ray detections were observed for this source promptly that displayed early flaring behavior at ${\sim}100$ days \footnote{For the remainder of the paper, days refers to days since discovery.} that was not well-explained \citep{guolo24-xrayfromopticalTDEs}. \citet{cendes24-mega} reported its first radio detection at late-times, 775 days post-discovery, with a previous upper limit observed at 126 days. This first detection was observed at 6.0 GHz with a flux density of $1.071 \pm 0.003$ mJy, corresponding to a luminosity of $\nu L_\nu=8.7 \times 10^{38}$ erg s$^{-1}$, with subsequent observations showing fading emission. The last reported 5.0 GHz observation places the flux density at $0.205 \pm 0.034$ mJy, or $1.7 \times 10^{38}$ erg s$^{-1}$, at 1262 days. The peak of this initial flare was constrained to be between 130-700 days after optical discovery from the earlier upper limits and the subsequent decay in the observed light curve. Initial outflow modeling estimated the outflow launch date at ${\sim}$ 400 days \citep{cendes24-mega}. \cite{alexander25-accretiondriven} plots the published X-ray and radio light curves simultaneously with estimated mass accretion rates to provide multi-wavelength context to the event. The light curves were compared to predictions of when a jet may be launched at super-Eddington and sub-Eddington accretion phases. It was found that the timing of the late time radio flare of AT2019ehz is generally consistent with a jet launched at peak accretion explanation.

AT2019eve was optically discovered on 2019 May 05 at $z$ = 0.081 \citep{vanVelzen21-tdediscs}. The initial discovery report in \cite{vanVelzen21-tdediscs} stopped short of a classification for AT2019eve noting the presence of HeI lines, a fast optical rise, and marginally lower blackbody temperatures. \cite{vanVelzen21-tdediscs} suggested a SN Ibn classification for AT2019eve as an explanation for the HeI lines; however they noted the presence of strong Balmer lines is inconsistent with such an explanation and left the classification as unknown. Later evidence of reddening after optical peak and only faint UV detections was reported in passing by \cite{hammerstein22-ztf}. The public data for AT2019eve is minimal, and no late-time host galaxy spectrum was ever made public, which complicates a definitive classification of this event. Despite its peculiarities, we believe a TDE is still the most likely explanation, especially with the context of its late time radio behavior, and it is often included within TDE samples (e.g. \citealt{cendes24-mega} \& \citealt{alexander25-accretiondriven} ). 

As reported in \citet{cendes24-mega}, AT2019eve was first detected at late-times in radio at 769 days post-discovery with a flux density of $0.766 \pm 0.009$ mJy at 6.0 GHz, or $\nu L_\nu=7.6 \times 10^{38}$ erg s$^{-1}$, after an upper limit obtained at 526 days post-optical discovery. Emission declined in each of the following observations with a final reported 5.0 GHz flux density $F_\nu=0.430 \pm 0.020$ mJy, or $\nu L_\nu=3.6 \times 10^{38}$ erg s$^{-1}$. Extrapolating the outflow radius across epochs to earlier times, the outflow was estimated to have been launched at ${\sim}545$ days post-discovery \citep{cendes24-mega}. The published multi-wavelength data for AT2019eve was also analyzed by \cite{alexander25-accretiondriven}. The radio flare was reported to be delayed with respect to the estimated sub-Eddington accretion phase by several hundred days, suggesting either incorrect accretion predictions or alternative mechanisms producing the radio emission.

\section{Observations} \label{sect:observations}

We obtained new radio observations of AT2019ehz and AT2019eve at multiple epochs between 2023 July 12 and 2025 October 03 using the Karl G. Jansky Very Large Array (VLA).  Details including dates and frequency ranges for each of these observation are provided in Table \ref{tab:observations}.

\begin{table}
    \centering
    \begin{tabular}{ccccc}\toprule
         Object&  Obs. Date&  $\delta$t&  Program& Frequency Band(s)\\\midrule
 AT2019ehz& 2023 July 22& 1545& 23A-241
(PI: Cendes)& L - X\\
 ...& 2023 September 12& 1597& SC240292
(PI: Hajela)& L - Ku\\
 ...& 2024 July 16& 1905& 24A-322
(PI: Cendes)& L - X\\
 ...& 2025 July 09& 2263& 
25A-159
(PI: Goodwin)& L - X\\
 ...& 2026 April 1& 2529& 26A-167
(PI: Christy)& L, X\\
 AT2019eve& 2023 September 15& 1594& 23A-241 
(PI: Cendes)& L - X\\
 ...& 2024 August 23& 1937& 24A-322 
(PI: Cendes)& L - X\\
 ...& 2025 April  16& 2173& 25A-159
(PI: Goodwin)& L - X\\
 ...& 2025 August 16& 2295& 25A-159
(PI: Goodwin)& C\\
 ...& 2025 October 03& 2343& 25B-154
(PI: Goodwin)&L, X\\ \bottomrule
    \end{tabular}
    \caption{Radio observation details for AT2019ehz and AT2019eve.}
    \label{tab:observations}
\end{table}

Data were processed using the standard procedures in the Common Astronomy Software Application package, CASA \citep{CASA}, and viewed with CARTA \citep{CARTA}. We used \verb|tclean| on the calibrated measurement sets downloaded from the NRAO archive and obtained all flux density measurements and uncertainties using \verb|imfit| and \verb|imstat|. When needed, further self-calibration was performed with \verb|auto_selfcal| developed by \citet{auto_selfcal}. An additional error of 10\% was added to the C- and X-band flux density measurements of AT2019eve taken at 1937 days to account for potential gain compression due to a thunderstorm over the array during the observation. All new flux measurements are provided in Table \ref{tab:fd-measurements} in the Appendix. We additionally use the flux density measurements from the two epochs of AT2019ehz and the three epochs of AT2019eve reported by \citet{cendes24-mega} to conduct our model fitting and analysis across all available epochs.

\section{Light Curves} \label{sect:light curve}

\subsection{Rebrightening TDE Sample 6 GHz Light Curve}

\begin{figure}[ht!]
\includegraphics[width=\columnwidth]{ 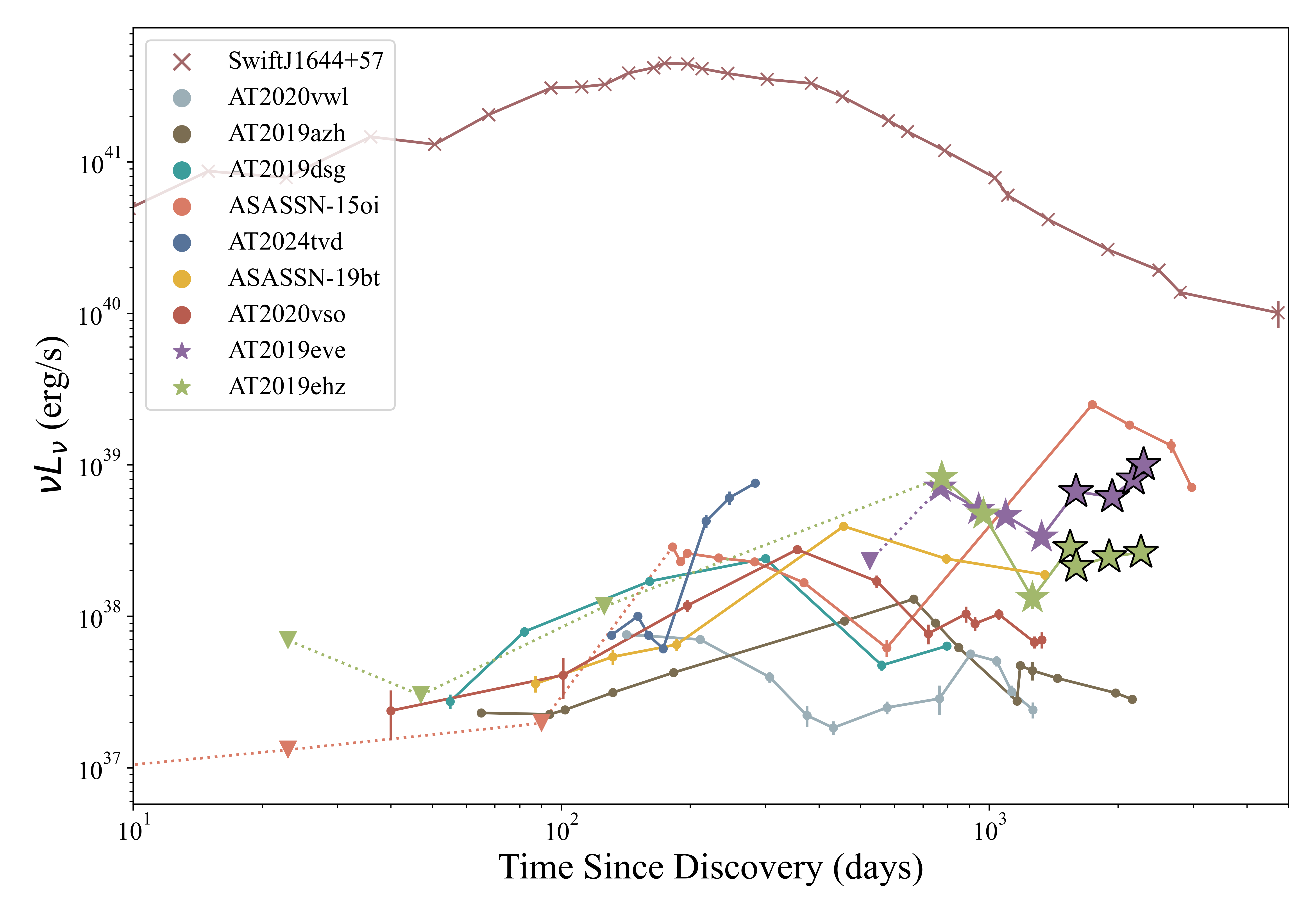}
\caption{Light curves at 5-6 GHz of all radio rebrightening TDEs to date in rest frame. Star markers used to denote AT2019ehz and AT2019eve. Stars with black outlines highlight new observations presented in this work. Data includes ASASSN-15oi \citep{hajela25-15oi}, AT2019ehz \citep[this work; ][]{cendes24-mega}, AT2019eve \citep{cendes24-mega}, ASASSN-19bt \citep{christy24-19bt}, AT2019dsg \citep{cendes24-mega}, AT2019azh \citep{goodwin22-19azh,burn25-19azh}, AT2020zso \citep{christy2025-20zso/21sdu}, AT2020vwl \citep{goodwin25-20vwl2},  and AT2024tvd \citep{sfaradi25-24tvd}. Sw J1644+57, a relativistic jetted TDE, is included for luminosity reference \citep{berger12-j1644-1,zauderer13-j1644-2,eftekhari18-j1644-3,cendes21-j1644-4}.
\label{fig:sample light curve}}
\end{figure}
Figure \ref{fig:sample light curve} presents the updated 5 GHz radio light curves of AT2019ehz and AT2019eve alongside previously reported TDEs that exhibit multiple episodes of radio flaring. For AT2019ehz, the rebrightening of emission was detected at 1545 days post-discovery with a 5.0 GHz flux density of $0.422 \pm 0.013$ mJy, corresponding to a rest-frame luminosity of $2.7 \times 10^{38}$ erg s$^{-1}$. This is an increase in emission which was previously $0.205 \pm 0.034$ mJy \citep{cendes24-mega}, and corresponds with a rise by a factor of ${\sim}2$. Subsequent observations show a slight decrease in flux followed by relatively steady emission for the next ${\sim}700$ days. 

For AT2019eve, rebrightening emission was first detected at 1594 days post-discovery with a flux density of $0.867 \pm 0.010$ mJy at 5.0 GHz, corresponding to a luminosity of $6.7 \times 10^{38}$ erg s$^{-1}$. This rise shows a factor of ${\sim}2$ increase in luminosity from the previous detection, which was a luminosity of $3.6 \times 10^{38}$ erg s$^{-1}$ \citep{cendes24-mega}. The luminosity is still rising with the last 5.0 GHz observation showing a flux density of $1.293 \pm 0.018$ mJy, corresponding to a luminosity of $1.01 \times 10^{39}$ erg s$^{-1}$.

\subsection{Individual Frequency Light Curves}

\begin{figure}[ht!]
\includegraphics[width=.49\linewidth]{ 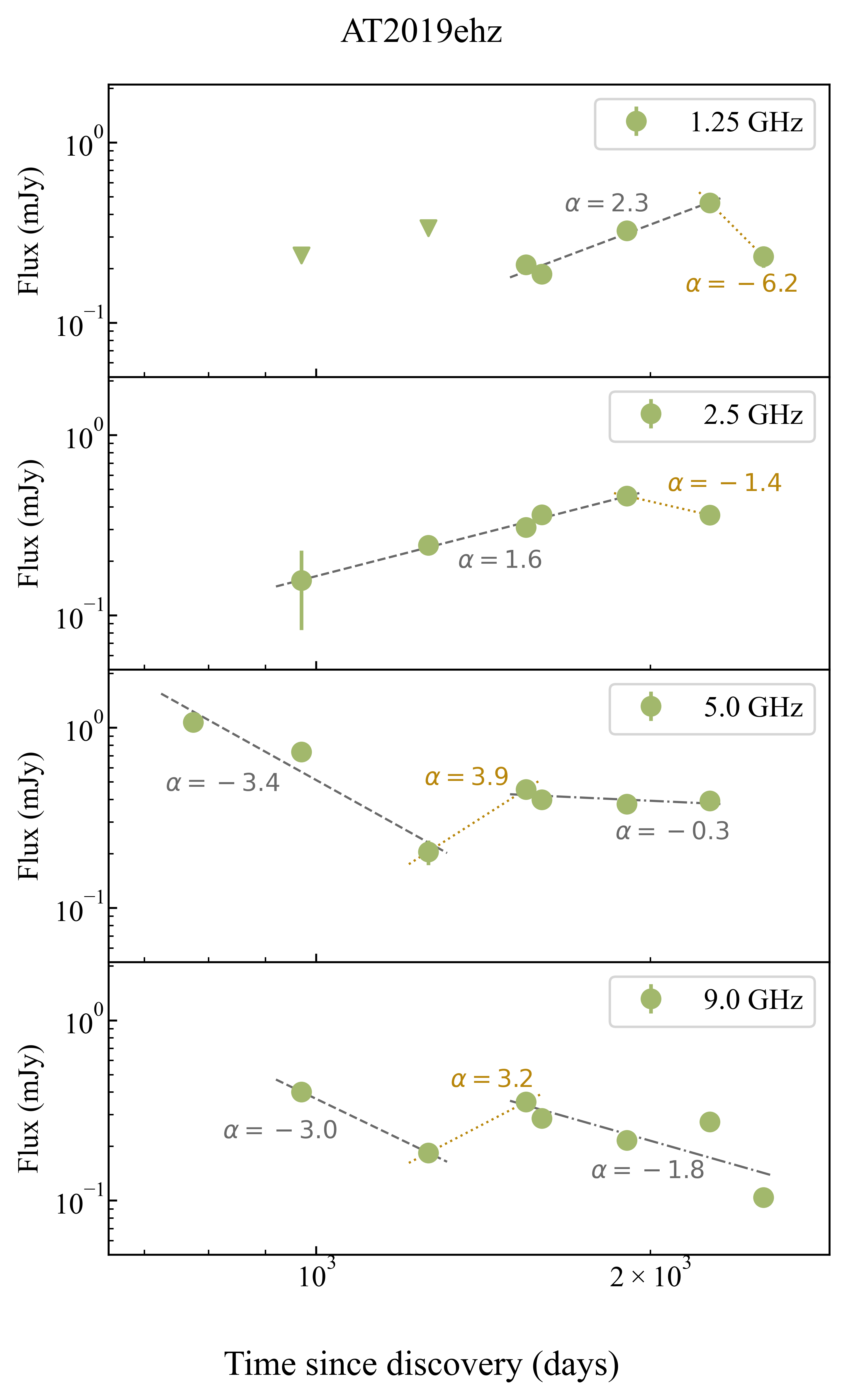}
\includegraphics[width=.49\linewidth]{ 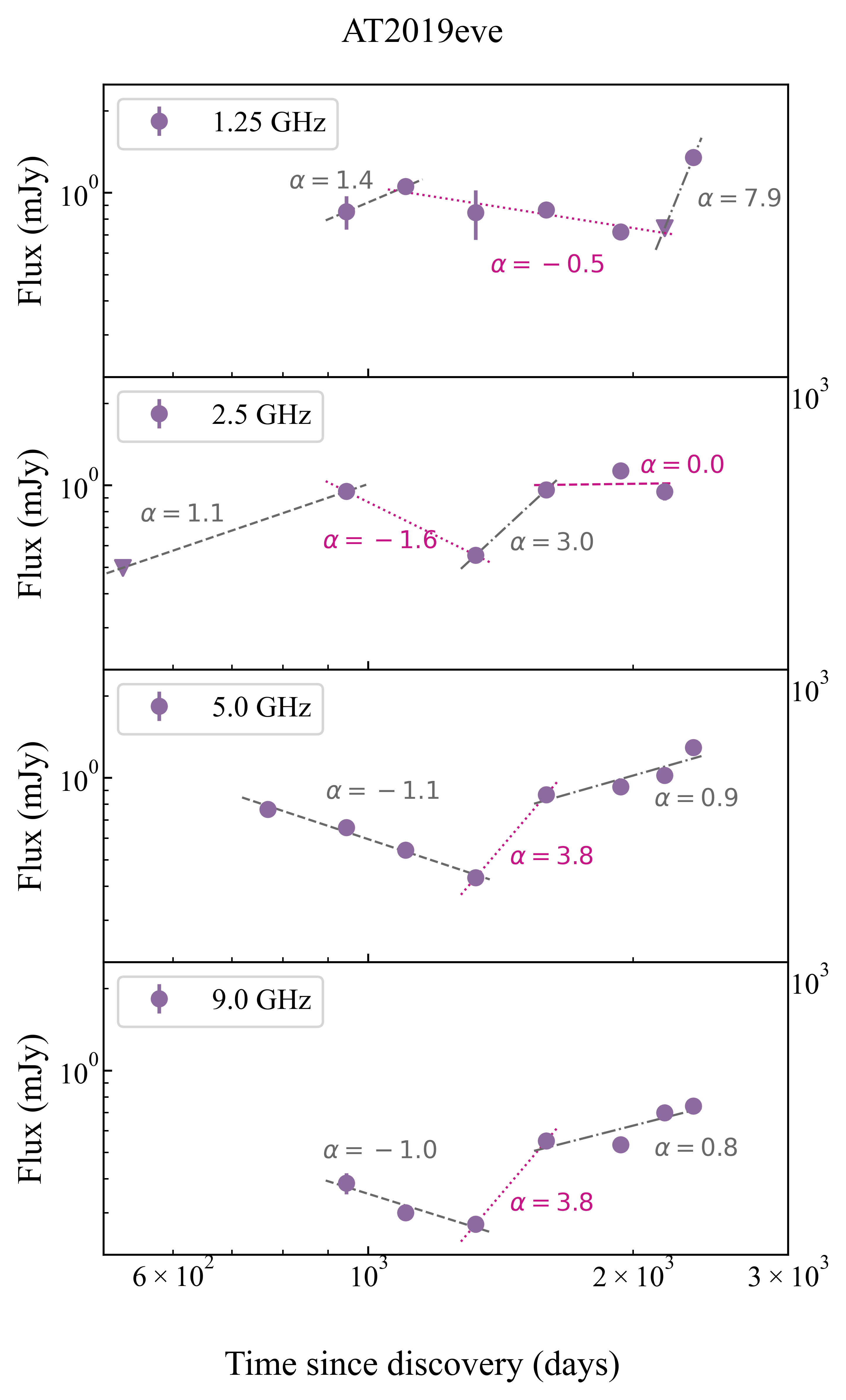}
\caption{Light curves of individual frequencies for AT2019ehz (left) and for AT2019eve (right) at frequencies 1.25, 2.5, 5.0, and 9.0 GHz. Power-law fits to rise and decay segments of each curve are plotted in dotted and dashed lines respectively. 
\label{fig:fitted-lightcurves}}
\end{figure}
Figure \ref{fig:fitted-lightcurves} displays the light curves of the lower halves of the frequency bands L-X (1.25, 2.5, 5.0, and 9.0 GHz) to investigate the rebrightening behavior across the spectrum. Simple power-law fits are shown to characterize rise and fall timescales for various segments of the light curve and $\alpha$ represents the slope of each fitted segment. For AT2019ehz (left), at 5.0 and 9.0 GHz, we see a rebrightening event where the flux increases at a rate of $\alpha >3$. For this TDE, the double peak component is not clearly evident in the lower frequency bands of this source at similar times. This behavior could be indicative of an additional synchrotron-radiating component superimposed on the pre-existing emission and evolving with time, as similarly seen in AT2020vwl, where the peak of the secondary flare at 11.0 GHZ is seen at least 250 days before the peak is seen at 1.25 GHz \citep{goodwin25-20vwl2}. 

In contrast, AT2019eve exhibits a rebrightening across all frequencies at similar times. The rise of this second flare of emission also increases at a rate of  $\alpha \ge3$ initially between 1325 and 1594 days after discovery and then remains roughly constant throughout the subsequent observations for 2.5--9.0 GHz. The exception is 1.25 GHz which shows the rebrightening event at 2343 days in which the flux rises at a rate of $\alpha >7.9$. The fit from the upper limit at 2173 to the detection at 2343 days serves as a lower limit to the rise as any flux value below the upper limit at 2173 would require an even steeper rise to produce the flux observed at 2343 days. This could be evidence of a new synchrotron component adding to the already observed emission, as discussed for AT2019ehz. The case for these additional components in either source can be more confidently examined once the secondary peaks of emission can be constrained.

Regardless of origin of emission, both AT2019ehz and AT2019eve display a rebrightening event associated with a second flare of emission around $\approx1550-1600$ days post-optical discovery.  This also corresponds with $\approx750-800$ days after initial radio detection, though we note the radio emission could have started much earlier for both TDEs due to a lack of early systematic monitoring. We have concluded the observed variations could not be explained by interstellar scintillation (see Appendix \ref{sect:scintillation}).

\section{Spectral Fitting} \label{sect:spectral fitting}
\begin{figure*}[ht!]
\includegraphics[width=\linewidth]{ 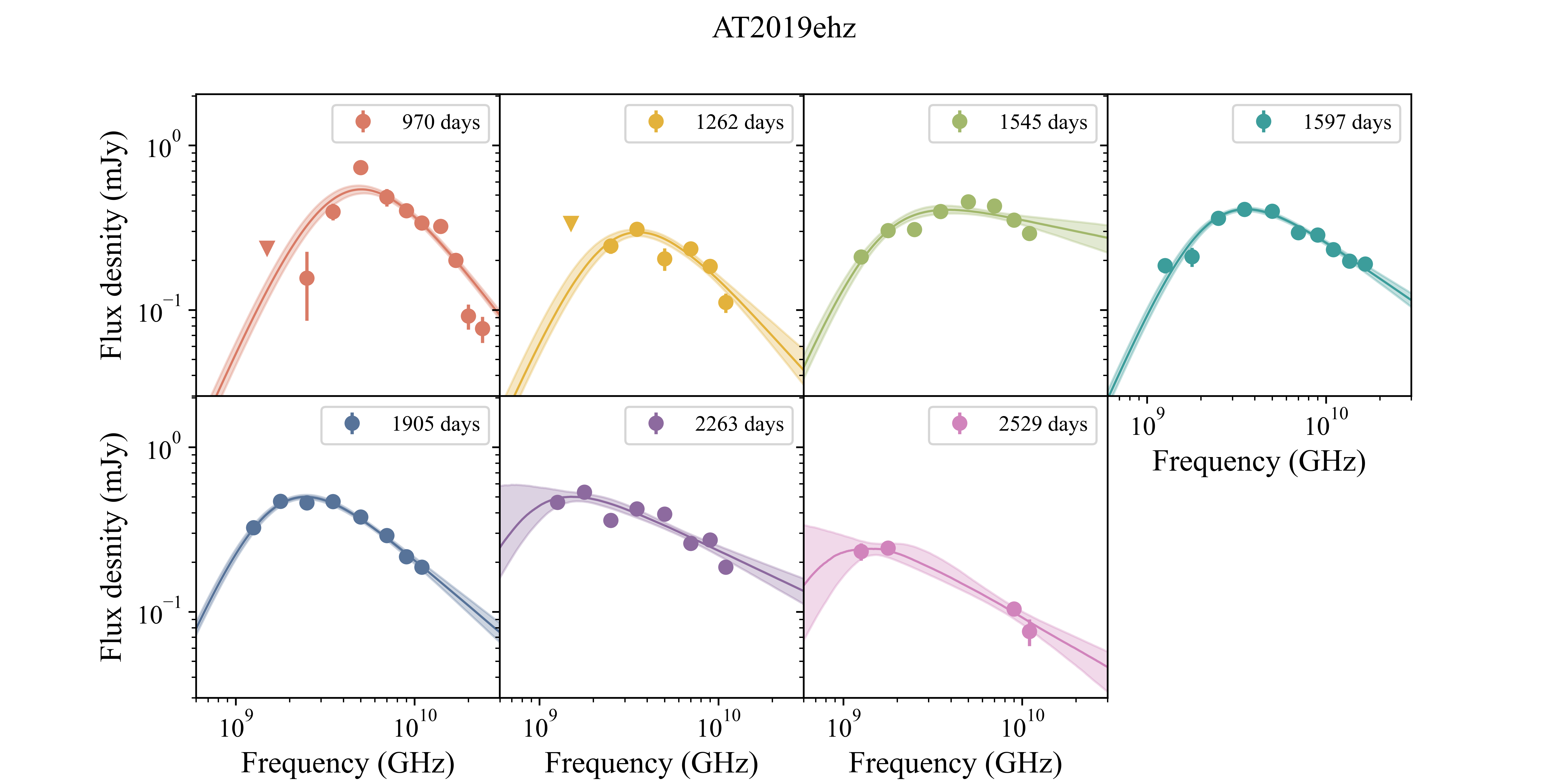}
\includegraphics[width=\linewidth]{ 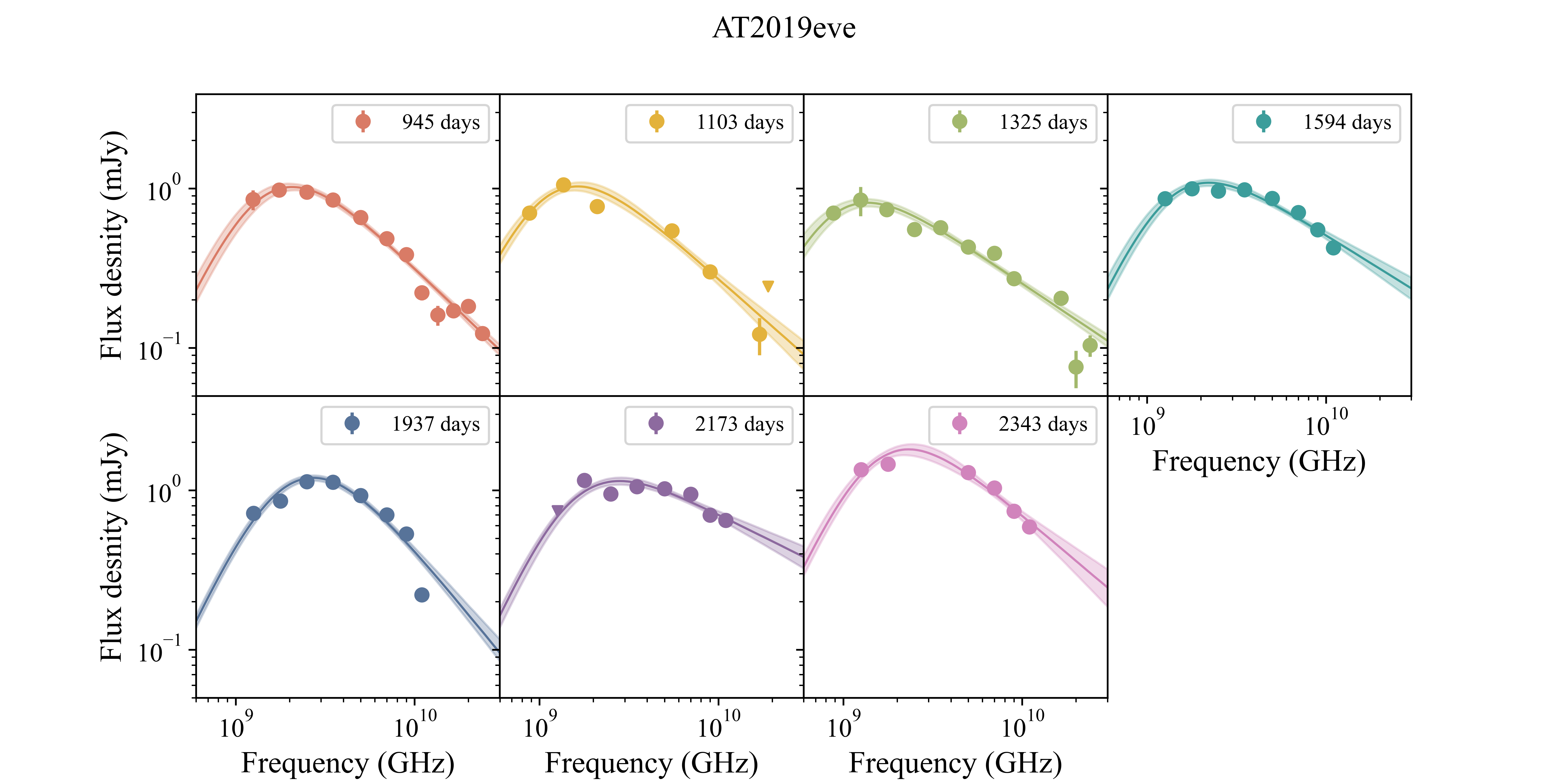}
\caption{Spectral energy distributions for AT2019ehz (top) and AT2019eve (bottom). Data for the first two epochs of AT2019ehz and the first three epochs of AT2019eve reported in \cite{cendes24-mega}. SED at 2319 days is constructed from combination of two observations from 2295 days and 2343 days. 5.0--11.0 GHz data points from 1937 days include additional error on measured flux density due to weather during observation.
\label{fig:SEDs}}
\end{figure*}
By combining the flux measurements from a single multi-frequency observation, spectral energy distributions (SEDs) can be constructed to characterize the radio emission. Radio emission from TDEs is commonly assumed to be dominated by synchrotron processes. In this context, the SEDs can be accurately modeled using the framework developed by \citet{granot02-spectral} for gamma-ray burst afterglows, which are also primarily governed by synchrotron emission. In this work, we assume non-relativistic emission, as is most consistent with the luminosity observed from the two sources being discussed. The model is given by:
\begin{equation} 
    F_{\nu}=F_{\text{ext}}\left[\left(\frac{\nu}{\nu_a}\right)^{-s\beta_1}+ \left(\frac{\nu}{\nu_a}\right)^{-s\beta_2}\right]^{-1/s}
\end{equation}
where $\nu_a$ is the synchrotron self-absorption frequency, $F_{ext}$ is the overall flux normalization, $\beta_{1}$ and $ \beta_{2}$ are the asymptotic spectral slopes above and below the peak (break 5 in \cite{granot02-spectral}), and $s$ defines the smoothing of the peak. Here, $\beta_{1} = 5/2 $, $\beta_{2} = (1-p)/2$, and $s=1.25-0.18p$, where $p$ is the power law index of the electron energy distribution. 

Fitting this model to all available SEDs of both TDEs, we estimate the best fit for the parameters $F_{ext}$, $\nu_a$, and $p$ using MCMC methods through \verb|emcee| \citep{emcee}. We assume the measurements have a Gaussian likelihood and use a logarithmic distribution sampling for priors of $F_{ext} = [.01 - 10]$ mJy and $\nu_a = [0.1 -11.0]$ GHz. For $p$, we assume a uniform prior of $p=[1.0-4.0]$. Furthermore, we include an additional parameter for systematic uncertainty beyond statistical uncertainty, $\sigma \lesssim 10\%$. The module was run using 200 MCMC chains for 10,000 steps with 2000 burn-in to achieve convergence based on an Effective Sample Size of $>300$ and a R-hat of $<1.05$. 
\begin{table}
    \centering
\caption{Spectral energy distribution best-fit parameters.}
\label{tab:SED-params}
    \begin{tabular}{cccccc}\toprule
 Object&  $\delta$t (d)& $\nu_p$ (GHz)& $F_{\nu,p}$ (mJy)& $p$&$\sigma$\\\midrule
         AT2019ehz&  970&  $5.04^{+0.26}_{-0.26}$&  $0.54^{+0.03}_{-0.03}$&  $3.90^{+0.14}_{-0.08}$& $9.53^{+0.72}_{-0.35}$\\
         ...&  1262&  $3.57^{+0.37}_{-0.27}$&  $0.30^{+0.01}_{-0.01}$&  $3.40^{+0.58}_{-0.41}$& $5.25^{+3.33}_{-3.37}$\\
         ...&  1545&  $3.82^{+0.29}_{-0.60}$&  $0.41^{+0.02}_{-0.02}$&  $1.48^{+0.24}_{-0.28}$& $8.98^{+1.17}_{-0.73}$\\
         ...&  1597&  $3.57^{+0.14}_{-0.14}$&  $0.41^{+0.01}_{-0.01}$&  $2.50^{+0.15}_{-0.16}$& $2.52^{+1.18}_{-2.80}$\\
         ...&  1905&  $2.46^{+0.09}_{-0.09}$&  $0.50^{+0.01}_{-0.02}$&  $2.86^{+0.17}_{-0.20}$& $4.36^{+1.76}_{-2.37}$\\
         ...&  2263&  $1.53^{+0.80}_{-0.34}$&  $0.52^{+0.03}_{-0.10}$&  $2.03^{+0.21}_{-0.24}$& $8.44^{+1.86}_{-1.12}$\\
 ...& 2529& $1.32^{+0.90}_{-0.67}$& $0.27^{+0.03}_{-0.10}$& $2.24^{+0.26}_{-0.59}$&$3.14^{+1.69}_{-3.66}$\\\midrule
         AT2019eve&  945&  $2.08^{+0.13}_{-0.17}$&  $1.02^{+0.05}_{-0.05}$&  $3.15^{+0.13}_{-0.12}$& $7.53^{+1.75}_{-2.87}$\\
         ...&  1103&  $1.62^{+0.13}_{-0.13}$&  $1.03^{+0.06}_{-0.06}$&  $2.97^{+0.29}_{-0.23}$& $8.06^{+1.39}_{-2.18}$\\
         ...&  1325&  $1.37^{+0.12}_{-0.13}$&  $0.82^{+0.05}_{-0.04}$&  $2.50^{+0.12}_{-0.11}$& $9.06^{+0.69}_{-1.25}$\\
         ...&  1594&  $2.21^{+0.14}_{-0.15}$&  $1.09^{+0.05}_{-0.05}$&  $2.40^{+0.22}_{-0.21}$& $7.97^{+1.34}_{-1.54}$\\
 ...& 1937& $2.73^{+0.10}_{-0.11}$& $1.20^{+0.05}_{-0.05}$& $3.74^{+0.19}_{-0.32}$&$9.42^{+0.43}_{-0.80}$\\
 ...& 2173& $2.78^{+0.16}_{-0.11}$& $1.14^{+0.06}_{-0.06}$& $2.11^{+0.24}_{-0.21}$&$8.21^{+1.24}_{-1.72}$\\
 ...& 2343& $2.31^{+0.17}_{-0.17}$& $1.81^{+0.14}_{-0.14}$& $2.91^{+0.44}_{-0.38}$&$8.31^{+1.16}_{-1.55}$\\ \bottomrule
    \end{tabular}

\end{table}

\begin{figure*}[ht!]
\includegraphics[width=\linewidth]{ 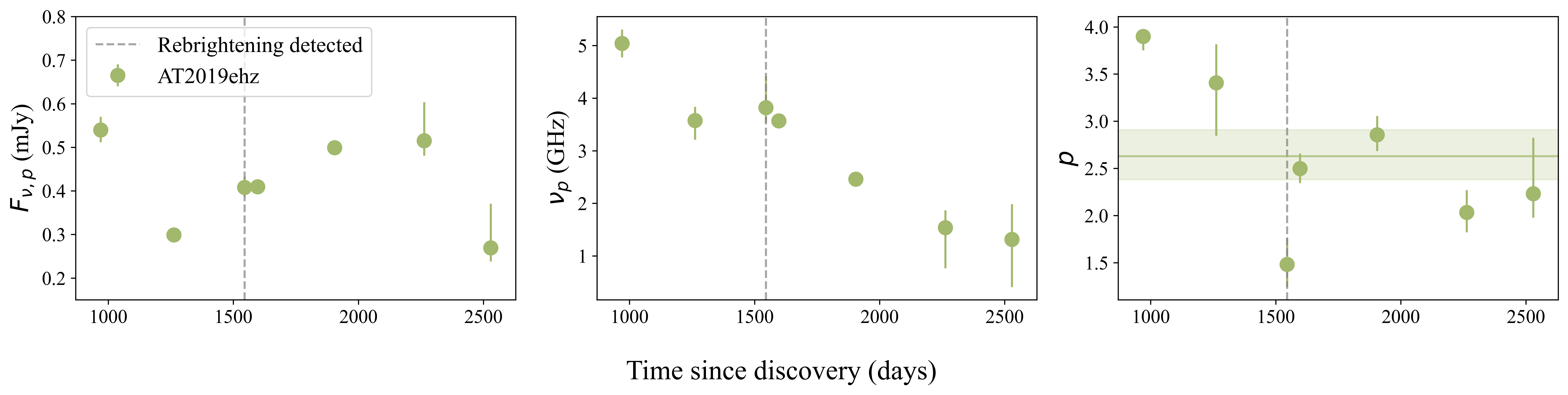}
\includegraphics[width=\linewidth]{ 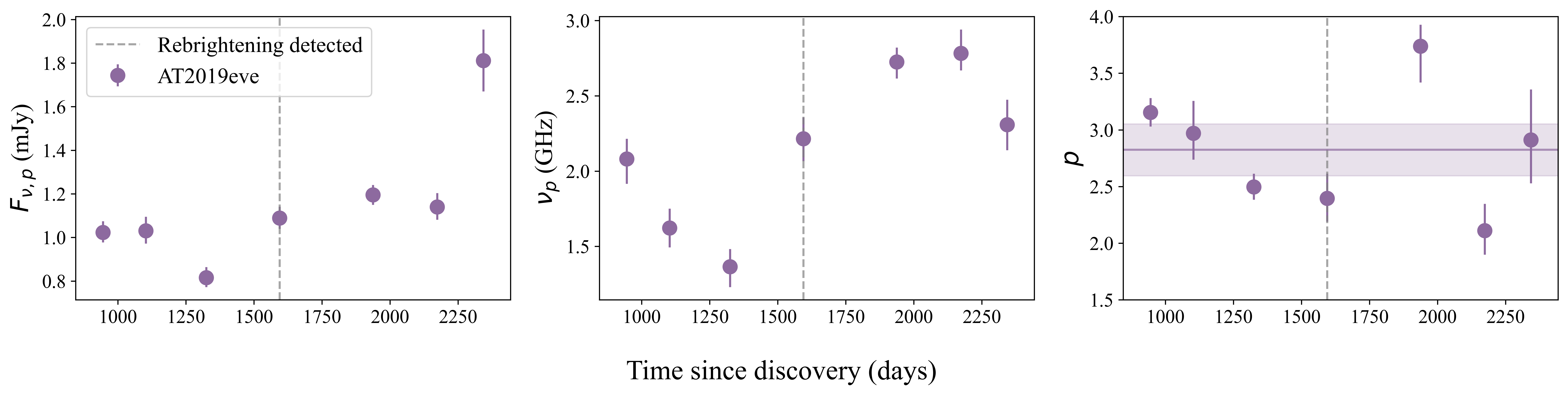}
\caption{Variation in flux of peak frequency ($F_{\nu,p}$), peak frequency ($\nu_p$), and electron distribution power law index ($p$) over time for AT2019ehz (top) and AT2019eve (bottom). Dashed gray line denotes when rebrightening in radio was initially detected. Solid horizontal lines in the right plots indicate the average p with the shaded region indicating the 1$\sigma$ error.
\label{fig:params-v-time}}
\end{figure*}

The fitted SEDs for AT2019ehz and AT2019eve are shown in Figure \ref{fig:SEDs}. The SEDs reported in \citet{cendes24-mega} were re-fit during this analysis to ensure consistency across the full dataset. The final SED of AT2019eve combines a C-band observation taken at 2295 days with L- and X-band observations taken at 2343 days. The best-fit parameters for each epoch are listed in Table \ref{tab:SED-params} and shown visually in Figure \ref{fig:params-v-time}. 

The SEDs of AT2019ehz show the peak frequency, $\nu_p$, decreasing from 5.4 GHz until the rebrightening in emission at 1545 days where $\nu_p$ increases slightly from 3.6 GHz to 3.8 GHz. The peak flux, $F_{\nu,p}$, increases by ${\sim}0.1$ mJy in the SED corresponding to the detection of the rebrightening and continues to rise. The fits exhibit a substantial spread in the inferred values of $p$ ($p = [1.48-3.90]$).  The lowest of these values, at 1545 days, may be consequence of the fitted model not matching the observed spectral shape well. We discuss possible explanations for this in Section \ref{sect:discussion}. We adopt a representative mean value of $p = 2.6$ for the remainder of the analysis to maintain consistency in modeling the outflows (see Section \ref{sect:outflow modeling}).

The SEDs of AT2019eve show that $\nu_p$ continually increases from 1.3 - 2.8 GHz from 1325 - 2173 days with a fairly consistent $F_{\nu,p}$ of ${\sim}1.1-1.2$ mJy after the rebrightening jump from 0.82 mJy to 1.09 mJy. The final epoch shows a substantial increase of $F_{\nu,p}$ by ${\sim}0.6$ mJy with a decrease in $\nu_p$ by ${\sim}0.3$ GHz. The value of $p$ at 1937 days is notably higher than in other SEDs, which we attribute to the higher uncertainty in flux in gain compression for this observation (see Section \ref{sect:observations}). Therefore, we adopt the mean value of $p = 2.8$ for the remainder of the analysis for consistency.

\section{Outflow Modeling} \label{sect:outflow modeling}
Using the best-fit parameters from Section \ref{sect:spectral fitting}, we can extract physical parameters of the observed outflow by modeling the radiation as synchrotron emission from a blast wave in which the peak of the spectrum constrains the size and minimal total energy of the emitting region \citep{barniolduran13-radius}. To do this we use an equipartition analysis where we assume the fraction of the postshock energy in the magnetic fields, $\epsilon_B$, is equal to the fraction in the relativistic electrons, $\epsilon_e$. Here, we set $\epsilon_b = \epsilon_e = 0.1$. The equipartition radius ($R_{eq}$) and kinetic energy ($E_{eq}$) are calculated using the following equations from \citet{barniolduran13-radius}, assuming a nonrelativistic spherical outflow:
\begin{equation}
\begin{split}
R_{\mathrm{eq}} \approx (1 \times 10^{17}~\mathrm{cm}) \times (21.8 \times 525^{p - 1})^{\frac{1}{13 + 2p}}   
\times \chi_e^{\frac{2 - p}{13 + 2p}} F_{p,\mathrm{mJy}}^{\frac{6 + p}{13 + 2p}} d_{L,28}^{\frac{2(p + 6)}{13 + 2p}} \nu_{p,10}^{-1} (1 + z)^{\frac{-(19 + 3p)}{13 + 2p}} \\
 \times f_A^{-\frac{5 + p}{13 + 2p}} f_V^{-\frac{1}{13 + 2p}}  
 4^{\frac{1}{13 + 2p}}  \epsilon^{1/17} \xi^{\frac{1}{13 + 2p}} 
 \end{split}
\end{equation}
\begin{equation}
\begin{split}
E_{\mathrm{eq}} \approx (1.3 \times 10^{48}~\mathrm{erg}) \times \left[21.8^{\frac{-2(p + 1)}{13 + 2p}}\right]  
\times \left[525^{p - 1} \chi_e^{2 - p} \right]^{\frac{11}{13 + 2p}} F_{p,\mathrm{mJy}}^{\frac{14 + 3p}{13 + 2p}} d_{L,28}^{\frac{2(3p + 14)}{13 + 2p}}  
\times \nu_{p,10}^{-1} (1 + z)^{\frac{-(27 + 5p)}{13 + 2p}} \\
\times f_A^{-\frac{3(p + 1)}{13 + 2p}} f_V^{\frac{2(p + 1)}{13 + 2p}} 4^{\frac{11}{13 + 2p}} \epsilon^{\frac{11}{13 + 2p}} \xi^{\frac{11}{13 + 2p}}  
 \times \left[\left(\frac{11}{17}\right) \epsilon^{-6/17} + \left(\frac{6}{17}\right) \epsilon^{11/17}\right]
 \end{split}
\end{equation}

Here, $d_L$ is the luminosity distance, $z$ is the redshift, $f_A=1$ and $f_V = 4/3 \times(1-0.9^3)=0.36$ are the area and volume filling factors, respectively, assuming the emitting region is a shell of thickness 0.1 $R_{eq}$, $\gamma_m = \text{max}[\chi_e(\Gamma-1),2]$ is the Lorentz factor for nonrelativistic sources, and $\chi_e = (p-2)(p-1)^{-1}\epsilon_e(m_p/m_e)$ where $m_p$ is the proton mass and $m_e$ is the electron mass. 

Using these calculated values, we can determine the following additional parameters: magnetic field strength, $B$,  Lorentz factor of electrons radiating at $\nu_a$, $\gamma_a$, and the number of radiating electrons, $N_e$ \citep{barniolduran13-radius}:
\begin{equation}
B \approx (1.3 \times 10^{-2}~\mathrm{G}) 
\times F_{p,\mathrm{mJy}}^{-2} d_{L,28}^{-4} (1 + z)^7 \nu_{p,10}^5 f_A^2 R_{\mathrm{eq},17}^4
\label{eq:B}
\end{equation}
\begin{equation}
\gamma_a \approx (525)  F_{p,\mathrm{mJy}} d_{L,28}^2 \nu_{p,10}^{-2}
\times (1 + z)^{-3} \frac{1}{f_A R_{\mathrm{eq},17}^2}
\label{eq:gamma_a}
\end{equation}
\begin{equation}
N_e \approx (4 \times 10^{54}) \times F_{p,\mathrm{mJy}}^3 d_{L,28}^6 (1 + z)^{-8}
\times f_A^{-2} R_{\mathrm{eq},17}^{-4} \nu_{p,10}^{-5} \left( \frac{\gamma_a}{\gamma_m} \right)^{(p - 1)}
\label{eq:Ne}
\end{equation}

\begin{figure*}[ht!]
\includegraphics[width=.49\linewidth]{ 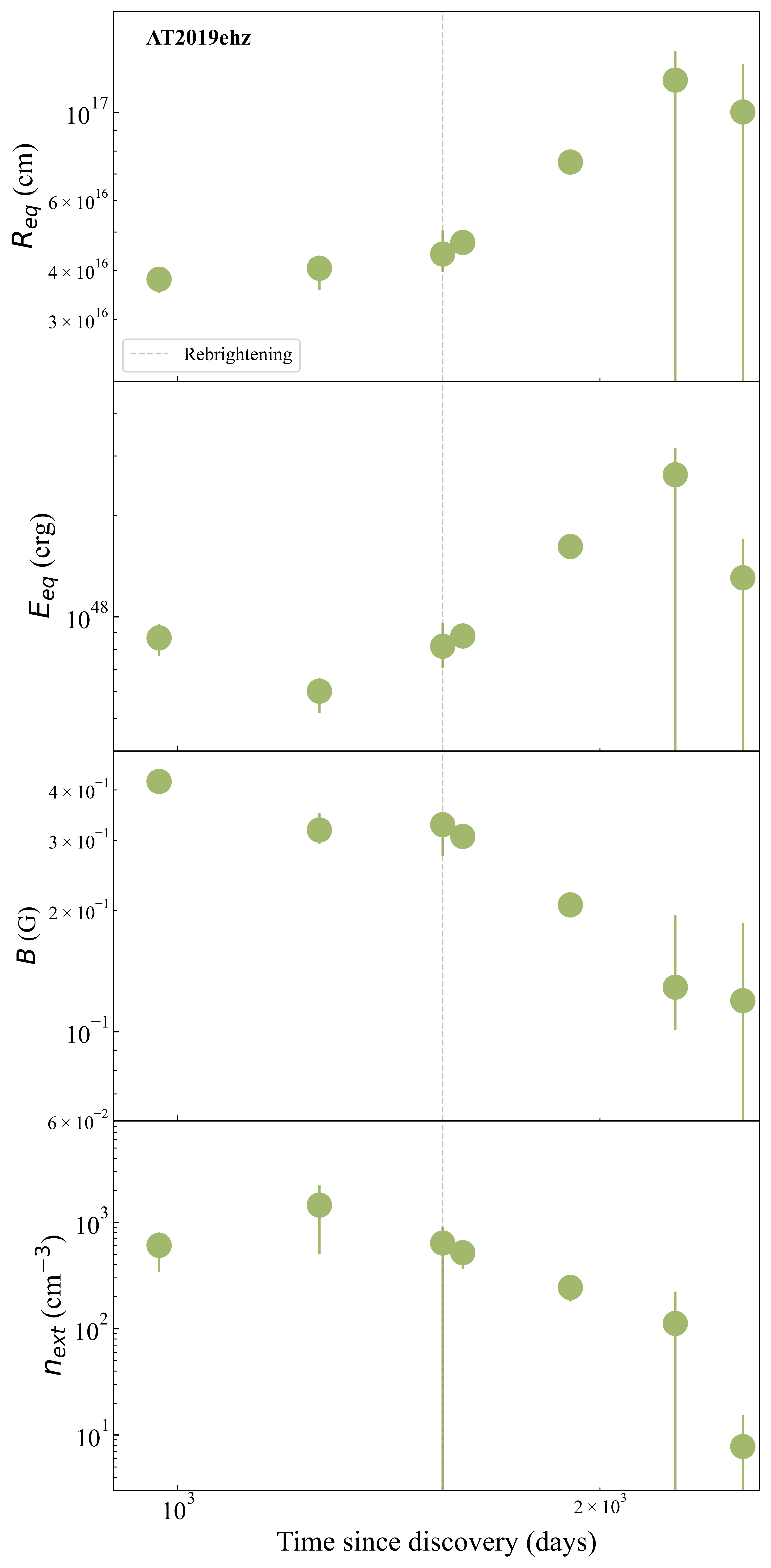}
\includegraphics[width=.49\linewidth]{ 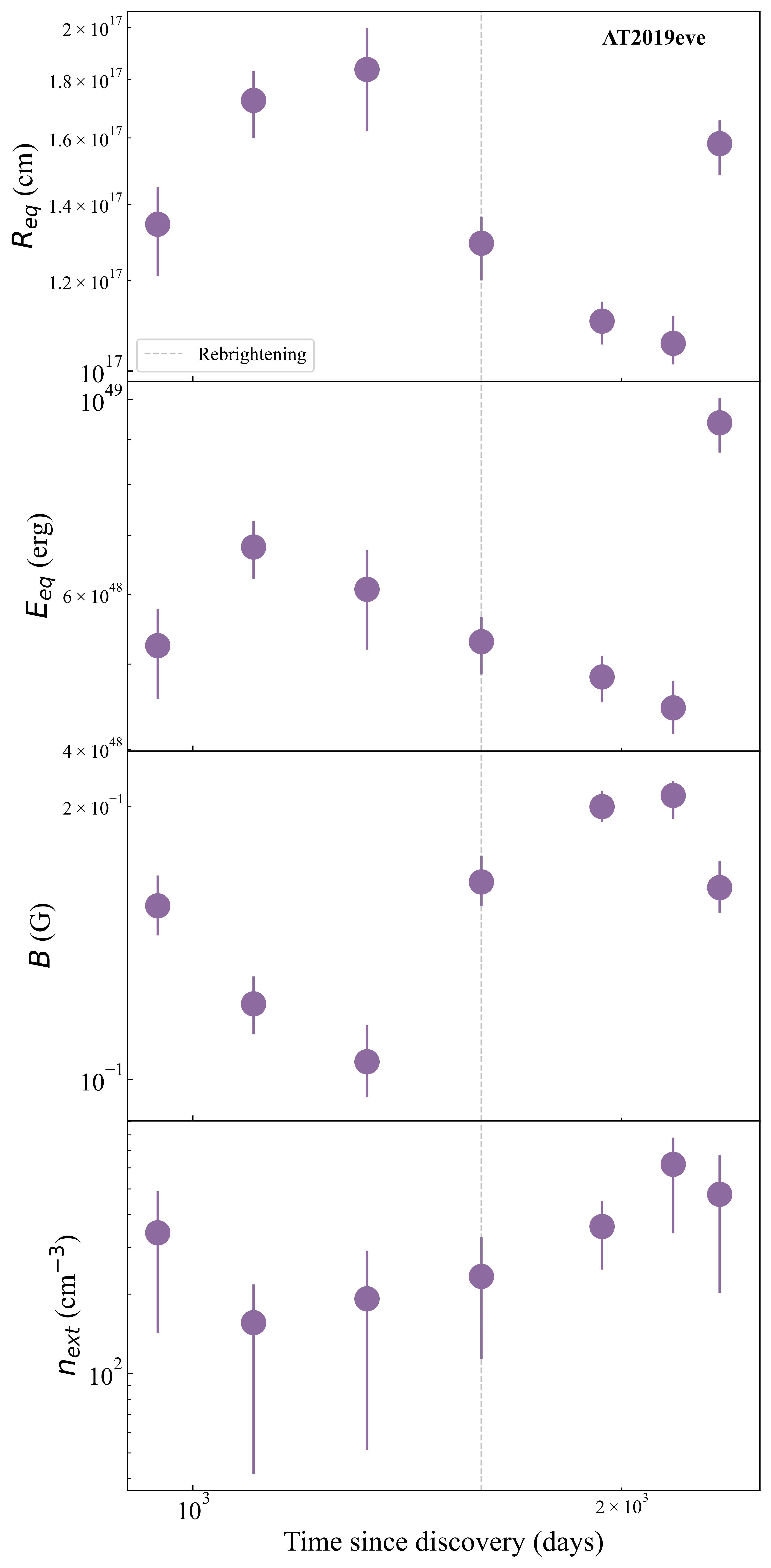}
\caption{Variation in equipartition radius ($R_{eq}$), equipartition energy ($E_{eq}$), magnetic field strength ($B$), and density ($n_{ext}$) over time for AT2019ehz (left) and AT2019eve (right). Dashed gray line denotes initial detection of rebrightening.
\label{fig:outflow params}}
\end{figure*}

The additional factor of 4 and the correction factor $\left( \frac{\gamma_a}{\gamma_m} \right)^{(p - 1)}$ on $N_e$  is for the nonrelativistic regime (Barniol Duran, private communication). We also calculate the density, $n_{ext}=N_e/4V$, where the volume of emission is $V=0.1R_{eq}$.  These results are summarized in Table \ref{tab:phys-params} and Figure \ref{fig:outflow params}. 
\begin{table}
    \centering
\caption{Equipartition model parameters.}
\label{tab:phys-params}
    \begin{tabular}{cccccccc}\toprule
         Object&  $\delta$t (d)&  log($R_{eq}$) (cm)&  log($E_{eq}$) (erg)&  log($B$) (G)&  log($N_{e}$)&  log($n_{ext}$) (cm$^{-3}$)& $\beta$ \footnote{$\beta\equiv\frac{R_{eq}}{ct}$ where t is days since estimated launch date as extrapolated for each flare (see Section \ref{sect:launch}).}\\\midrule
 AT2019ehz& 970& $16.58^{+0.03}_{-0.03}$& $47.94^{+0.05}_{-0.05}$& $-0.38^{+0.03}_{-0.02}$& $53.33^{+0.08}_{-0.07}$& $2.95^{+0.17}_{-0.16}$&$0.02^{+0.02}_{-0.01}$\\
 ...& 1262& $16.61^{+0.04}_{-0.05}$& $47.78^{+0.04}_{-0.06}$& $-0.50^{+0.05}_{-0.03}$& $53.18^{+0.19}_{-0.12}$& $2.71^{+0.33}_{-0.22}$&$0.01^{+0.01}_{-0.01}$\\
 ...& 1545& $16.64^{+0.07}_{-0.04}$& $47.91^{+0.08}_{-0.06}$& $-0.48^{+0.04}_{-0.07}$& $53.33^{+0.13}_{-0.23}$& $2.76^{+0.24}_{-0.45}$&$0.03^{+0.04}_{-0.03}$\\
 ...& 1598& $16.67^{+0.02}_{-0.02}$& $47.94^{+0.02}_{-0.02}$& $-0.51^{+0.02}_{-0.02}$& $53.21^{+0.08}_{-0.07}$& $2.55^{+0.12}_{-0.11}$&$0.03^{+0.03}_{-0.03}$\\
         ...&  1905&  $16.87^{+0.02}_{-0.02}$&  $48.21^{+0.02}_{-0.02}$&  $-0.68^{+0.02}_{-0.02}$&  $53.60^{+0.07}_{-0.07}$&  $2.34^{+0.13}_{-0.10}$& $0.03^{+0.03}_{-0.03}$\\
         ...&  2263&  $17.09^{+0.09}_{-0.36}$&  $48.43^{+0.10}_{-0.42}$&  $-0.89^{+0.33}_{-0.09}$&  $53.29^{+1.20}_{-0.34}$&  $1.41^{+2.34}_{-0.60}$& $0.04^{+0.05}_{-0.02}$\\
 ...& 2529& $17.00^{+0.17}_{-0.41}$& $48.11^{+0.15}_{-0.50}$& $-0.92^{+0.36}_{-0.17}$& $52.18^{+1.16}_{-0.74}$& $0.53^{+2.42}_{-1.25}$&$0.03^{+0.04}_{-0.01}$\\\midrule
         AT2019eve&  945&  $17.13^{+0.04}_{-0.04}$&  $48.72^{+0.05}_{-0.05}$&  $-0.81^{+0.04}_{-0.03}$&  $54.55^{+0.13}_{-0.11}$&  $2.53^{+0.25}_{-0.20}$& $0.05^{+0.11}_{-0.10}$\\
         ...&  1103&  $17.24^{+0.03}_{-0.03}$&  $48.83^{+0.03}_{-0.03}$&  $-0.92^{+0.03}_{-0.03}$&  $54.55^{+0.15}_{-0.15}$&  $2.19^{+0.22}_{-0.24}$& $0.06^{+0.11}_{-0.10}$\\
         ...&  1325&  $17.26^{+0.04}_{-0.05}$&  $48.78^{+0.05}_{-0.06}$&  $-0.98^{+0.05}_{-0.04}$&  $54.71^{+0.17}_{-0.14}$&  $2.28^{+0.32}_{-0.24}$& $0.05^{+0.09}_{-0.07}$\\
         ...&  1594&  $17.11^{+0.02}_{-0.03}$&  $48.72^{+0.03}_{-0.03}$&  $-0.78^{+0.03}_{-0.03}$&  $54.34^{+0.13}_{-0.11}$&  $2.37^{+0.23}_{-0.18}$& $0.03^{+0.05}_{-0.04}$\\
         ...&  1937&  $17.04^{+0.02}_{-0.02}$&  $48.68^{+0.03}_{-0.03}$&  $-0.70^{+0.02}_{-0.02}$&  $54.33^{+0.07}_{-0.06}$&  $2.56^{+0.13}_{-0.12}$& $0.02^{+0.03}_{-0.03}$\\
         ...&  2173&  $17.02^{+0.03}_{-0.02}$&  $48.65^{+0.03}_{-0.03}$&  $-0.69^{+0.02}_{-0.02}$&  $54.50^{+0.09}_{-0.10}$&  $2.79^{+0.13}_{-0.16}$& $0.02^{+0.03}_{-0.02}$\\
         ...&  2343&  $17.20^{+0.02}_{-0.03}$&  $48.97^{+0.03}_{-0.03}$&  $-0.79^{+0.03}_{-0.03}$&  $54.92^{+0.17}_{-0.14}$&  $2.68^{+0.23}_{-0.20}$& $0.03^{+0.03}_{-0.03}$\\\bottomrule
    \end{tabular}

\end{table}

Figure \ref{fig:outflow params} presents the best-fit parameters for $R_{eq}$, $E_{eq}$, $B$, and $n_{ext}$ for each epoch of AT2019ehz and AT2019eve. AT2019ehz shows a slower radial expansion prior to the detected rebrightening, traversing a total range ${\sim}0.5\times10^{16}$ cm in 575 days. The rate of radial traverse then increases drastically, moving ${\sim}6\times10^{16}$ cm in 984 days. The equipartition energy follows a similar trend of decreasing energy from ${\sim}10^{48}$ erg before the rebrightening and a steady increase in energy after, ranging from ${\sim}8\times10^{47}-3\times10^{48}$ erg. The magnetic field strength and density appear to consistently decrease across all epochs, with magnetic field strength declining from $0.4 - 0.12$ G and density decreasing from $900 - 3$ cm$^{-3}$. 

For AT2019eve, the radius appears to be \textit{decreasing} over a range of ${\sim}8\times10^{16}$ cm over a span of ${\sim}1000$ days. This evolution is likely artificial due to the model's dependence on the expectation of synchrotron emission to have a decreasing $\nu_p$ over time, which is not the case for the observed behavior of AT2019eve. Possible explanations for the source's deviation from the expected synchrotron evolution are discussed in Section \ref{sect:discussion}. The equipartition energy is seen roughly constant at $6\times10^{48}$ erg, with the exception of the latest observation where $E_{eq} = 1\times10^{49}$ erg. The magnetic field strength and density exhibit an inverse trend to the radial expansion, varying by approximately 0.1 G and 450 cm$^{-3}$, respectively, over the same 1000-day interval.

\subsection{Launch Date Estimation}
\label{sect:launch}
\begin{figure*}[ht!]
\includegraphics[width=.49\linewidth]{ 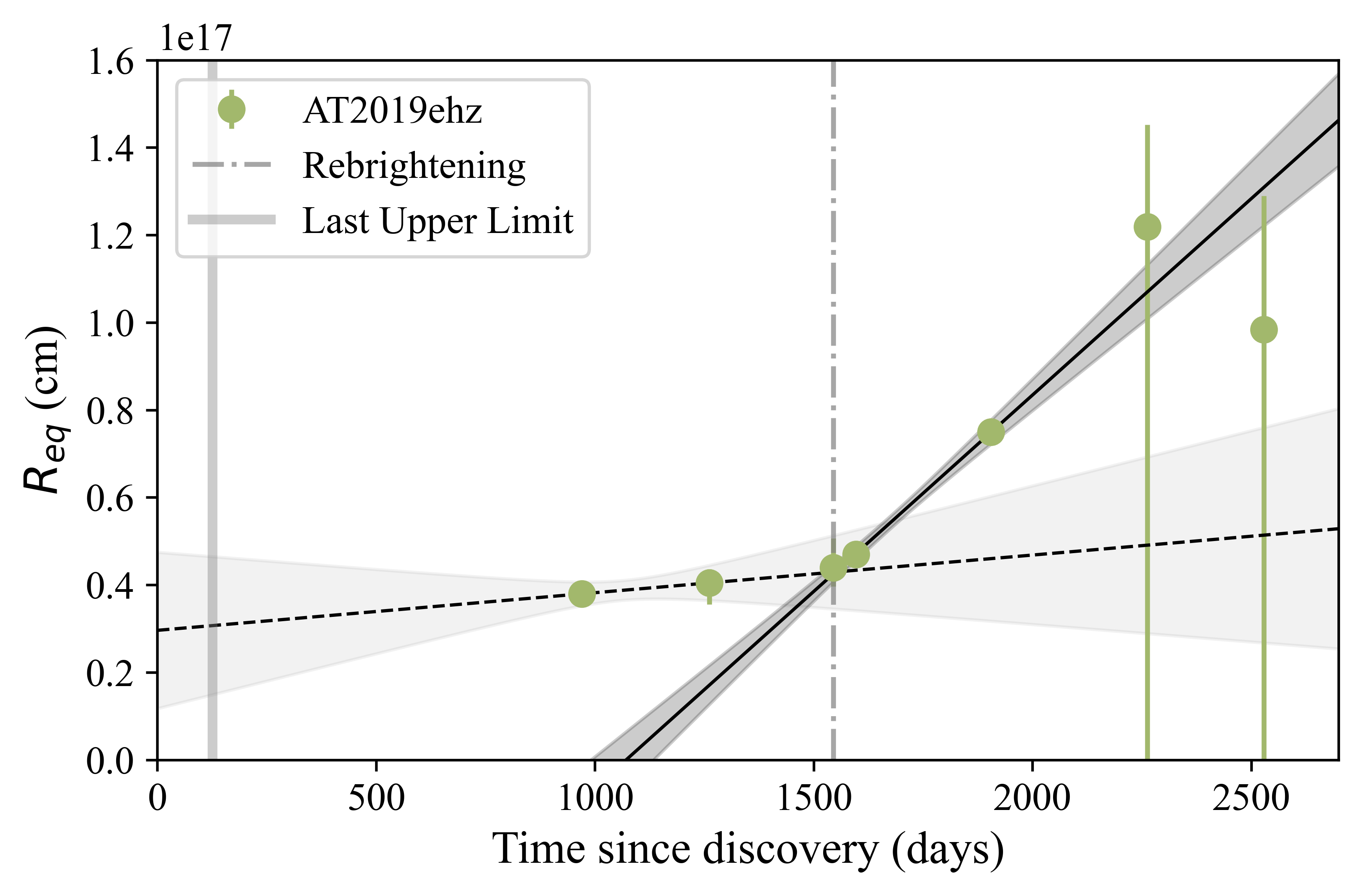}
\includegraphics[width=.49\linewidth]{ 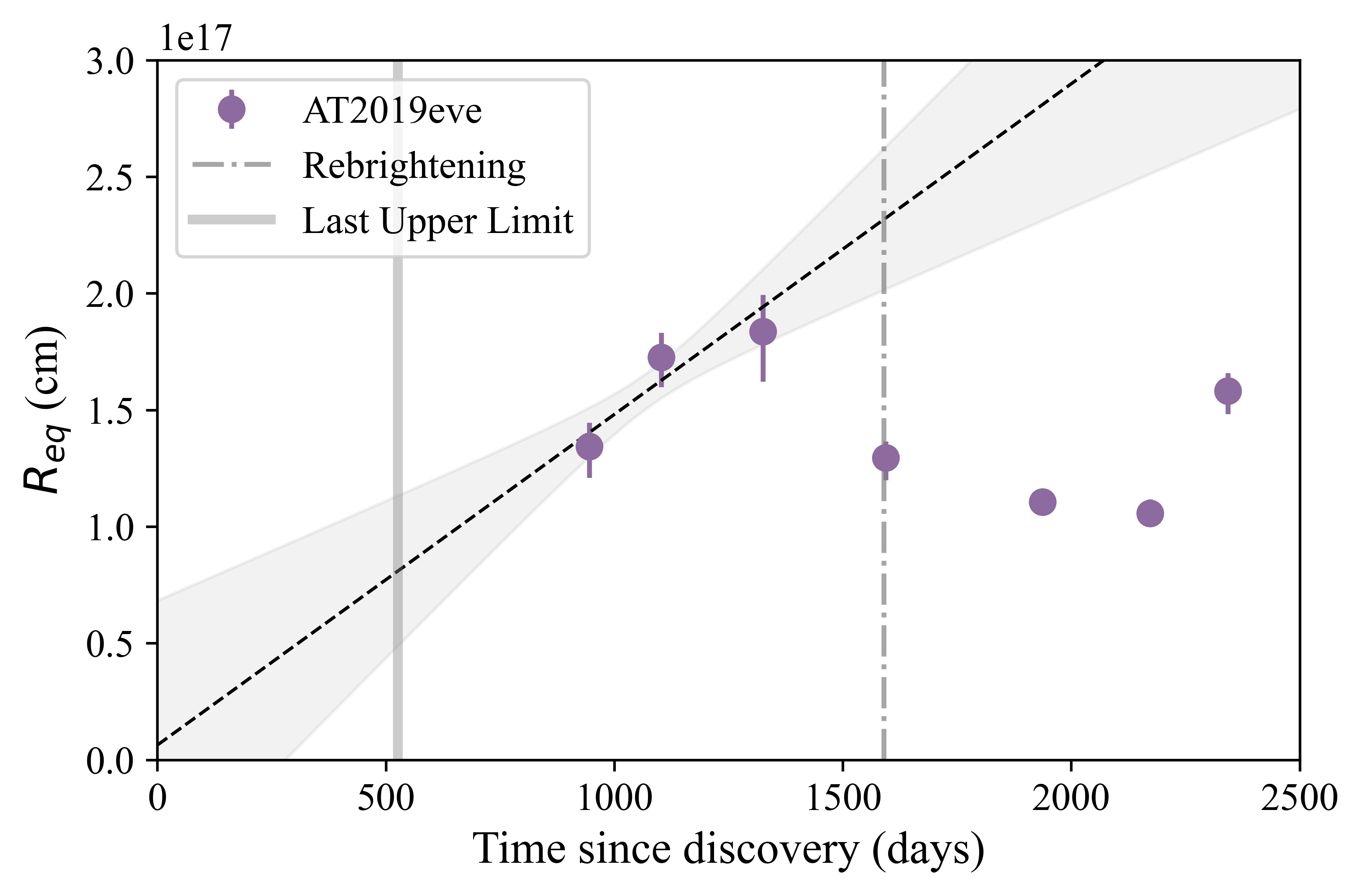}
\caption{Outflow radius over time for the two TDEs. Vertical lines mark the last upper limit prior to radio detection. \textit{Left:} The extrapolation for AT2019ehz, where a dotted line represents best-fit line of the first flare and solid line denotes that of the second flare. \textit{Right:} The extrapolation for AT2019eve, where the dotted line shows approximate initial expansion of the first flare. Shaded regions are 1$\sigma$ uncertainties.
\label{fig:radial expansions}}
\end{figure*}

Given multiple SEDs for a source over time, the evolution of $R_{eq}$ can be fit with a linear trend to estimate the outflow launch date, as seen in Figure \ref{fig:radial expansions}, and calculate the mean expansion velocity, $\bar\beta=v/c$, using the slope of the fitted line as $v$. Figure \ref{fig:radial expansions} shows the radial expansion fits for both AT2019ehz and AT2019eve. 

The first flare of AT2019ehz does not converge to a launch date, as is consistent with \cite{cendes24-mega}, thought to be due to the outflow already decelerating at time of observations. The launch date was estimated by \cite{cendes24-mega} to be at ${\sim}400$ days when assuming similar rise times to observed fall times. At time of observation, the first flare of AT2019ehz appears to be moving at $\bar\beta = 0.003$ until rebrightening where the radial expansion rate drastically increases to $\bar\beta = 0.033$. This extrapolates back to a second outflow launch date of $1071^{+62}_{-79}$ days. 

Due to the atypical behavior of AT2019eve and the apparent recession of the inferred equipartition radius, the expansion fit over this temporal interval does not converge to a well-defined launch date.  This indicates a more complex environment than contained by this model. Nevertheless, we estimate that the initial outflow evolution appears to move at $\bar{\beta} = 0.05$ before the decreasing behavior takes place. The fit to the first data points give a launch date of the initial outflow. The extrapolation of the epochs prior to rebrightening results in a launch date of $-45^{+324}_{-763}$ days for the first flare. Given the last upper limit prior to radio detection, a non-detection at 3 GHz at 526 days \citep{cendes24-mega}, we find it likely that the true launch would occur on the later side of the estimated range at $\sim250$ day. This launch window is inconsistent with the previously estimated launch date of 545 days, found by \cite{cendes24-mega}, because the downward trend seen at 1325 days was initially thought to be evidence of a change in outflow structure due to deceleration and thus was left out of the fit for that flare. As the new observations of AT2019eve provide additional context that no longer suggests a decelerating outflow scenario, we include this point in the expansion fit of the first flare, resulting in a slower expansion and earlier launch estimation.

\section{Discussion} \label{sect:discussion}

To date, several theoretical frameworks have been proposed to explain scenarios that would produce delayed radio emission associated with TDEs.  To analyze the origins of the secondary radio flares reported here for AT2019ehz and AT2019eve, we compile the previously reported properties of the flare and physical outflow, examine each of these as possible origins of the rebrightening of the objects discussed in this work, and then compare them to the sample of TDEs reported to have rebrightened at late-times.

\subsection{Origins of radio rebrightening}
The observed and modeled radio behavior of AT2019ehz is suggestive of two distinct outflows with extrapolated launch dates of ${\sim}400$ \citep{cendes24-mega} and ${\sim}1070$ days (see Section \ref{sect:outflow modeling}), respectively. This interpretation is supported by the observed accelerated expansion where $\bar\beta\simeq0.003$ increases to $\simeq0.033$ after the rebrightening event (Figure \ref{fig:radial expansions}). It is further supported by the steady increase in kinetic energy from $6\times10^{47}- 2.7\times10^{48}$ erg, pointing to a second outflow which is more energetic and launched at higher speeds. Additionally, the SED at 1545 days, where the rebrightening was first detected, also exhibits the lowest electron energy distribution of $p=$1.5.  This may be the result of multiple superimposed spectra producing a more shallow spectral index in the optically thin segment of the SED, as was seen in ASASSN-15oi \citep{hajela25-15oi}.

In contrast, current efforts to model the observed behavior of AT2019eve do not yet favor any specific outflow interpretation. The generally consistent spectral shape and inferred kinetic energy across all epochs of the SEDs suggests a single population of synchrotron-emitting electrons with no clear evidence of a new outflow. However, the steadily increasing $\nu_p$ from 1.4 to 2.8 GHz across SEDs ranging over ${\sim}850$ days is unusual behavior that sets it apart from the expected evolution of synchrotron emission usually seen from TDEs. Typical models of synchrotron emission outflows are thus limited in this situation as they assume $\nu_p$ to be steadily decreasing with outflow evolution \citep[i.e., ][]{granot02-spectral}. These limitations may thus be the explanation of the apparent decreasing outflow radius.  This could be due to a period of superimposed emission from separate synchrotron-emitting regions that artificially drove a steady increase in $\nu_p$.  

However, we note the most recent SED of AT2019eve, though composed of two separate observations almost 50 days apart, shows an increase in peak frequency and decrease in $\nu_p$, as expected in our outflow models (see Figure \ref{fig:params-v-time}). The flux of the peak frequency in this SED also increases by over 0.6 mJy. Further monitoring of the radio evolution of AT2019eve will allow us to conclude if this is the beginning of a transition in its radio evolution or simply an artifact of the combined SED.  New models exploring multiple outflows of various geometries and at various launch times could be helpful in constraining the origin of the unexpected evolution of AT2019eve.

\subsubsection{Interactions with variable circumstellar densities}

Shock interactions with varying density profiles in the circumnuclear medium (CNM) can induce fluctuations in radio luminosity, analogous to those observed in certain supernovae (e.g. SN2007bg; \citealt{salas13-sn2007bg}). \citet{matsumoto24-bondi} propose that at the Bondi radius, the CNM density profile, characterized by $n \propto R^{-k}$, may undergo a transition from a steep $k = 2$ profile to a flatter $k = 0$ regime. They demonstrate that shock interactions occurring at this transition would result in the optically thin flux to evolve as $\propto t^3$. This theory is also black hole temperature dependent, which is not understood. For more dramatic radio luminosity increases, an alternative variable-density environment has been proposed, in which the CNM is structured by molecular clouds in the surrounding SMBH environment, as discussed by \citet{zhuang25-clouds}. In this scenario, interactions between the outflow and these dense molecular clouds can produce significantly steeper rises in the radio emission, with the characteristic rise time being dependent on the width of the outflow or density enhancement.

At the time of the second flare for AT2019ehz, the optically thin portion of the spectrum was $>4$ GHz. Thus, we can examine the 5.0 and 9.0 GHz light curves, as seen in Figure \ref{fig:fitted-lightcurves}.  The rise index of the second flare is $\alpha >3$ for this both 5 and 9 GHz (Section \ref{sect:light curve}), which is too steep to be explainable by the Bondi radius scenario. Further, this rise rate is consistent with the molecular cloud models, but the increasing equipartition energy seen after the rebrightening event is not well-explained by this explanation. We therefore conclude that the delayed secondary flare of AT2019ehz is likely not a consequence of interactions with an environment of varying density.

The optically thin portion of the spectrum for AT2019eve during the second flare was $>3$ GHz, and therefore we can also utilize the fitted 5.0 and 9.0 light curves, where the rise index was seen to be $\alpha {\sim}3.8$ for this part of the spectrum (Figure \ref{fig:fitted-lightcurves}). Similar to AT2019ehz, this is too steep to be explainable by the Bondi radius scenario, but not too steep for a molecular cloud origin model. We find the latter to be a potential explanation for the observed increase in flux density, especially with the reasonably stable energy across the flares seen from this object. If the density is a complex environment, simple models may not be sufficient to describe certain observed behavior such as the peak frequency evolution.

\subsubsection{Decelerating off-axis jet}
There is a critical apparent velocity, $\beta_{eq,N}<0.44$ where $\beta_{eq,N}\equiv\frac{(1+z)R_{eq,N}}{ct}$, in which it is proposed that a nonrelativistic spherical outflow and a relativistic off-axis jet are both valid solutions to observed delayed radio flares \citep{matsumoto23-offaxis,beniamini23-offaxis}. In the relativistic off-axis jet case, emission would become visible as the jet decelerates and the emitting region widens into a viewable angle (as has been suggested for AT2018hyz; \citealt{cendes22-18hyz,sfaradi24-18hyz,cendes26-18hyz_continued}). As the jet continues to decelerate and the entire emitting region becomes visible, $\beta_{eq,N}$ is expected to surpass that critical velocity threshold, making the off-axis jet solution viable only if that threshold is reached.

The velocities for the first and second outflow of AT2019ehz were inferred to be $\bar\beta\simeq0.003$ and $0.033$, respectively, which would corresponds to $\beta_{eq,N}\simeq0.003$ and 0.035 (Section \ref{sect:outflow modeling}). Both these velocities are below the $\beta_{eq,N}=0.44$  threshold, but the velocity appears to be consistent across the flare evolution, making it unlikely that speeds will ever increase enough to reach the critical threshold necessary of an off-axis jet scenario. Additionally, the flare has kinetic energies of ${\sim}10^{48}$ erg with luminosities ${\sim}10^{38}$ erg s$^{-1}$, which are both consistent with other nonrelativistic spherical outflows rather than expected behavior from off-axis relativistic jets. This suggests that an off-axis jet is an unlikely interpretation for the origin of the AT2019ehz secondary flare. A similar conclusion was reached regarding the possible origin of the first delayed flare from \cite{cendes24-mega}.

As the apparent reverse motion of AT2019eve renders the velocity fit to the expansion curve unreliable, we can examine the apparent velocities calculated at each epoch with respect to the optical discovery date. The $\beta$, listed in Table \ref{tab:phys-params}, is at a maximum during the initial flare at ${\sim}0.06$, corresponding to a $\beta_{eq,N}=0.065$, but continues to decrease across epochs to $\beta=0.03$ at the most recent observation. The declining nature of the outflow velocity indicates that the apparent velocity of this source also will likely not reach the needed $\beta_{eq,N}=0.44$. The lower luminosities of ${\sim}10^{38}$ erg s$^{-1}$ and energies of ${\sim}10^{49}$ erg are also inconsistent with an off-axis jet scenario, similar to AT2019ehz. However, we note our values for $\beta$ for AT2019eve are difficult to measure accurately, due to the decreasing radii measured in our model.  While further monitoring of this source would be beneficial before an off-axis jet origin can be entirely ruled out for AT2019eve, the speeds of the initial outflow alone are not likely to reach the the needed $\beta_{eq,N}$.

\subsubsection{Delayed accretion driven outflows}
Observations and modeling have shown that delayed outflows may be driven by activity in accretion evolution at times of delayed super-Eddington accretion as well as at sub-Eddington when the accretion rate transitions below a critical rate of ${\sim}0.02-0.03 \dot{M}_{edd}$ \citep{alexander25-accretiondriven,hajela25-15oi,goodwin25-20vwl2}. This latter behavior mirrors a phenomenon seen in X-ray binaries when accretion disks transition from a soft to hard state \citep{fender04-xraybinaryjets}.  This points to a potential scale-invariant critical accretion rate at which compact jets are launched, as recently proposed by \cite{goodwin26-universalaccretion}. 

An accretion-induced wind or jet created at peak accretion was the most favored scenario for the first delayed flare of AT2019ehz by \cite{cendes24-mega}. \cite{alexander25-accretiondriven} modeled the optical flare for AT2019ehz using \verb|MOSFiT| \citep{mosfit}, and found the transition across an accretion rate of ${\sim}0.03\dot{M}_{edd}$ would coincide with an outflow launch date of ${\sim}300$ days, which is generally consistent with the launch date of ${\sim}400$ estimated by \citet{cendes24-mega} using outflow modeling. This indicates the first delayed flare may originate from an accretion state change. The accretion rate is predicted to continue in a steady decline, placing the accretion rate well below the critical sub-Eddington accretion rate at the estimated time of the second outflow launch. However, these predictions are likely underestimates of the accretion rate, as they are modeled with the fallback rate rather than the viscous timescale that the accretion disk evolves on \citep{alexander25-accretiondriven}. Therefore, these predicted accretion rates better serve as a lower limit on expected accretion rate at these later times. 

A more robust estimation could be inferred using methods based on the viscous timescale, like those seen in \cite{goodwin26-universalaccretion}. Unfortunately, due to lack of disk emission detections in optical/UV at late times, the methods used in \cite{goodwin26-universalaccretion} are unable to constrain disk and black hole parameters for either AT2019ehz or AT2019eve. Thus, with the accretion rates estimated by \cite{alexander25-accretiondriven} as a lower bound for AT2019ehz at these late times, it is possible that the first radio flare originates from a period of super-Eddington accretion, and the second is the a result of a state transition.

For AT2019eve, the first flare was not detected until several hundred days after the accretion rate should have transitioned across the critical sub-Eddington threshold rate of  ${\sim}0.03\dot{M}_{edd}$ \citet{alexander25-accretiondriven}. This implies either a delay in accretion relative to predictions, or other origins are needed to account for the observed radio flare. As is the case for AT2019ehz, the findings in \citet{alexander25-accretiondriven} are likely underestimating the accretion rate at these times due to model assumptions. Thus, a much more delayed accretion evolution than that predicted by \cite{alexander25-accretiondriven} is possible, allowing the flares to be coincident with these two states of expected outflow-producing accretion. Additionally, the consistent radio spectral shape across epochs for AT2019eve is not indicative of two independent outflows that would contribute multiple components within the spectra (e.g. seen in AT2020zso \citealt{christy2025-20zso/21sdu}), making a separate outflow scenario less likely.

\subsubsection{Repeating partial TDEs}
It is suspected that a number of TDEs are in fact partial TDEs (pTDEs), where the stellar core survives the initial disruption \citep{zhong22-pTDErates}. In specific cases, the stellar remnant can be placed on an orbital trajectory that leads to one or more subsequent close passages to the SMBH.  When this happens the remnant may undergo repeated pTDEs, continuing until the stellar mass is fully disrupted or the orbital path no longer permits further disruption. There have been reported cases of repeating pTDEs, each exhibiting new optical/X-ray flares with each disruption (e.g. \citealt{liu23-pTDEeRASSt,hinkle24-22dbloptical,makrygianni2025-22dblstandardpTDE}). In addition to a re-flare of optical/X-ray light, it can be assumed that an outflow generating radio emission could be produced following each disruption event.

AT2019ehz did exhibit a very dim second optical flare ${\sim}300$ days after the initial optical flare \citep{langis26-repeating}. This second optical flare is approximately 1200 days prior to the second radio flare. To date, no optical/UV/X-ray rebrightening has been detected for AT2019eve. Lack of consistent sampling across all frequencies for either source leads to no evidence of connection between secondary optical flares and their observed secondary radio flares, but the paucity of data means we cannot definitively rule out this scenario.

\subsection{Comparison to Sample} 
To date, including this work, there have been 9 reported TDEs that have have exhibited multiple flares in their radio light curves: ASASSN-15oi \citep{hajela25-15oi}, AT2019ehz (this work, \citealt{cendes24-mega}), AT2019eve (this work, \citealt{cendes24-mega}), ASASSN-19bt \citep{christy24-19bt}, AT2019dsg \citep{cendes24-mega}, AT2019azh\footnote{AT2019azh was reported to have a rebrightening event at 15.5 GHz by \citet{sfaradi22-19azh}, however similar features were not ever observed in lower frequencies \citep{goodwin22-19azh,burn25-19azh}. As a result, we include the source in the sample for the light curve seen in Figure \ref{fig:sample light curve} for completeness, but exclude it from the sample for the remaining analysis.} \citep{sfaradi22-19azh, goodwin22-19azh}, AT2020zso \citep{christy2025-20zso/21sdu}, AT2020vwl \citep{goodwin25-20vwl2},  and AT2024tvd \citep{sfaradi25-24tvd}. Their light curves can be seen in Figure \ref{fig:sample light curve}. These previously reported double-peaked radio TDEs each exhibit a distinct prompt flare followed by a rebrightening late-time flare. AT2019ehz and AT2019eve are the only two TDEs in this sample in which both distinct flares are observed $\gtrsim2$ years after optical discovery, but the existence of an undetected prompt flare cannot be entirely ruled out due to paucity of data in early times for these sources. 

For each of these TDEs, two separately launched outflows are possible explanations of the observed flares, where the initial flare is near the time of disruption ($<100$ days), and the second outflow launch date varies widely across the sample (${\sim}100 - 1000$ days after optical discovery). For example, AT2020zso has a second outflow launch estimate of around 850 days after optical discovery, making the outflow launch separation approximately 800 days \citep{christy2025-20zso/21sdu}. AT2024tvd, on the other hand, was estimated to launch its secondary outflow just 100 days post-disruption after the initial outflow \citep{sfaradi25-24tvd}.  The separation of launch dates for the potential outflows in AT2019ehz are ${\sim}600$ days, but the initial flare was at  ${\sim}$400 days (see Section \ref{sect:outflow modeling}). For AT2019ehz, the last non-detection was at 126 days post-discovery followed by an initial radio detection at 776 days, so the possibility of a prompt flare detected within this window cannot be ruled out. For AT2019eve, the only upper limit prior to initial radio detection at 769 days post-discovery was at 526 days, also leaving open the possibility of a missed prompt flare. Almost all of the rebrightening TDEs exhibit a brief jump up in $\nu_p$ near the time of their associated secondary flares, though the true duration of this observed increase is limited by the cadence of observations. The only other TDE to date that has been observed to increase in $\nu_p$ across observation is AT2020vwl, which showed an increase in $\nu_p\propto t^3$ for ${\sim}$500 days. AT2019ehz, in comparison, has a much longer duration of this increasing $\nu_p$ behavior than the rest of the sample.

\begin{figure}[ht!]
\centering
\includegraphics[width=.8\linewidth]{ 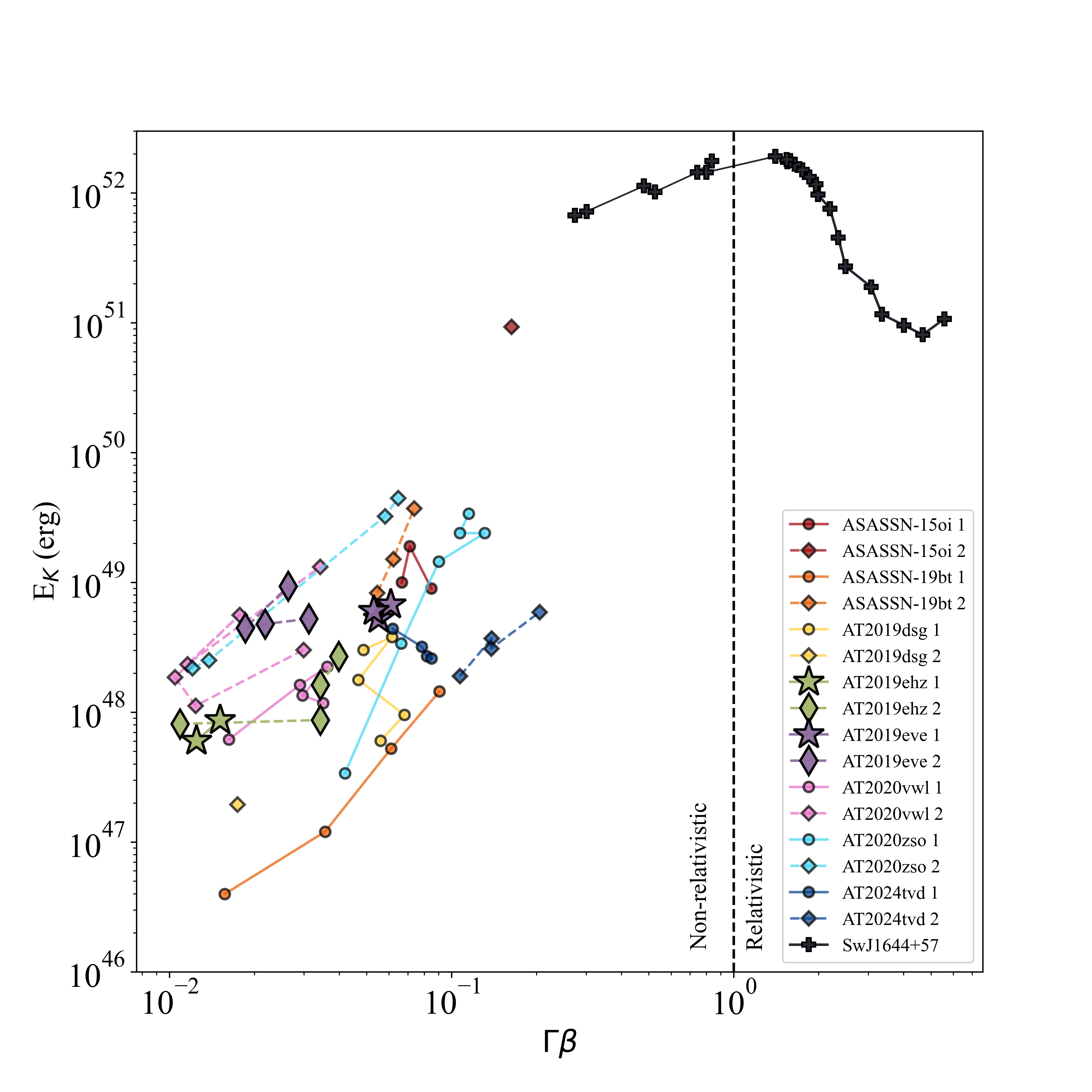}
\caption{The kinetic energy versus velocity for AT2019ehz and AT2019eve from data presented in this work, along with other rebrightening TDEs reported to date. Circles and solid lines indicate data from the first flare, with diamonds and dashed lines indicating data from second flares. We also include the relativistic TDE Sw J1644+57 for reference, and the vertical dashed line denotes the relativistic velocity boundary. Data includes ASASSN-15oi \citep{hajela25-15oi}, AT2019ehz \citep{cendes24-mega} (this work), AT2019eve \citep{cendes24-mega} (this work), ASASSN-19bt \citep{christy24-19bt}, AT2019dsg \citep{cendes24-mega},  AT2020zso \citep{christy2025-20zso/21sdu}, AT2020vwl \citep{goodwin25-20vwl2},  and AT2024tvd \citep{sfaradi25-24tvd}, and Sw J1644+57 \citep{berger12-j1644-1,zauderer13-j1644-2,eftekhari18-j1644-3,cendes21-j1644-4}.
\label{fig:energy-vel}}
\end{figure}

We can also compare the physical properties of TDEs with rebrightenings, both with each other and in terms of their first and second flares.  In Figure \ref{fig:energy-vel}, we show the kinetic energies and velocities for all reported TDEs with radio light curves that include rebrightening episodes. Data from the first flare are shown with a circle and solid connecting lines, and data from the second flare are denoted with a diamond and dashed connecting lines. We find that the energies and velocities inferred for AT2019ehz, AT2019eve, and the remaining rebrightening TDEs in our sample are broadly consistent with those observed in non-relativistic TDEs, which represent the dominant class of the TDE population \citep{alexander20-radprops}. Energies typically span from ${\sim}10^{48-49}$ erg across the sample with velocities ranging from less than $0.01c$, as seen in AT2019dsg \citep{cendes24-mega}, up to $0.5c$, as seen in the most delayed outflow launch estimation of AT2024tvd \citep{sfaradi25-24tvd}. AT2019ehz and AT2019eve are of average energies when compared to these TDEs, and appear to have typical velocities for this TDE population.  We see AT2019ehz increase in energy after the onset of the secondary flare which is also seen in the energy trends of AT2024tvd, AT2020vwl, and ASASSN-19bt \citep{sfaradi25-24tvd,christy2025-20zso/21sdu,goodwin25-20vwl2}. AT2019ehz has a generally consistent speed of $0.03c$ during the secondary flare. AT2019eve's speed is harder to estimate due to the model limitations, but individual epochs show velocities ranging from $0.02-0.1c$, with the velocity decreasing over time (Table \ref{tab:phys-params}).
\begin{figure}[ht!]
\centering
\includegraphics[width=.8\linewidth]{ 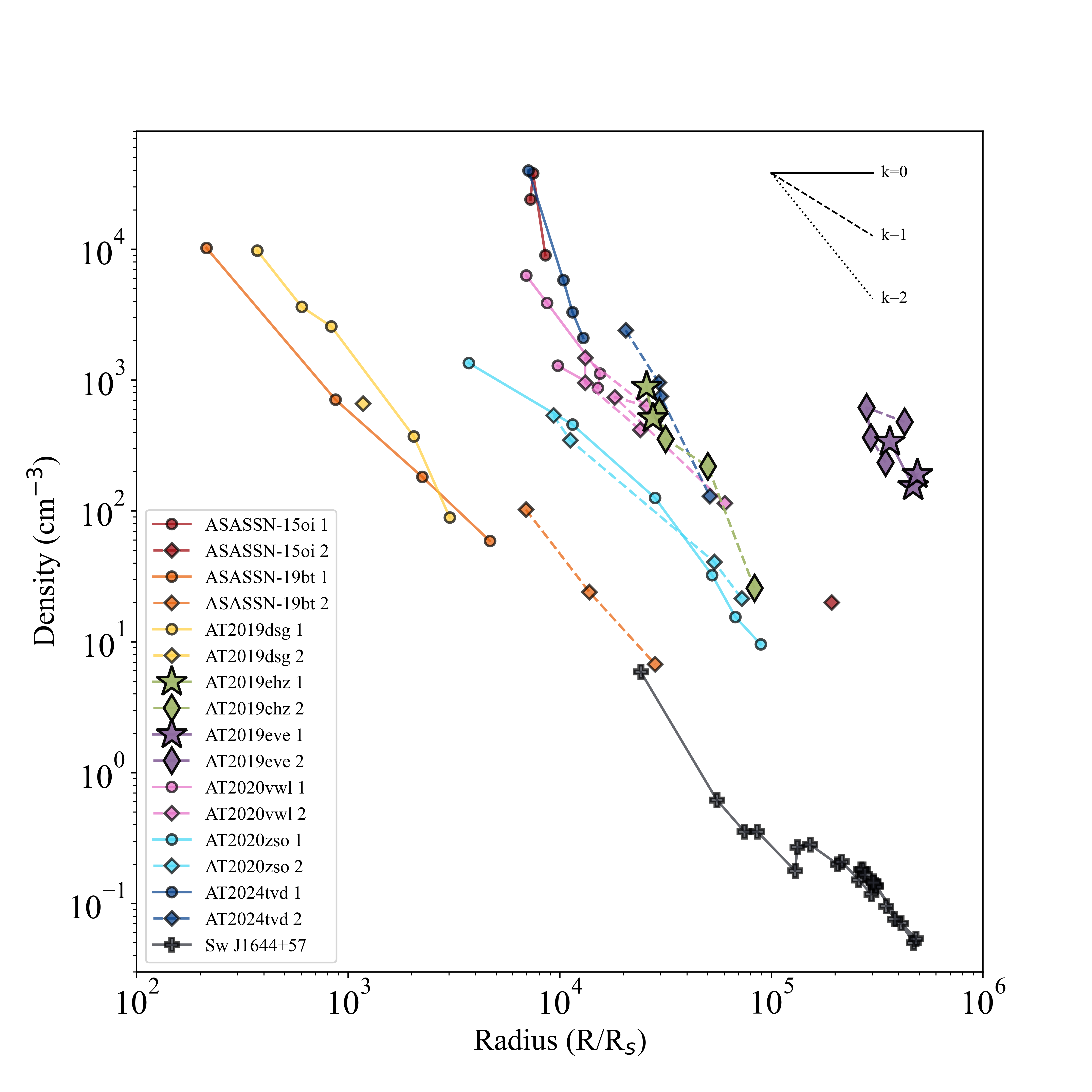}

\caption{The CNM density profiles for the TDEs AT2019ehz and AT2019eve, along with the rest of the rebrightening TDE sample, normalized to the Schwarzschild radius of each SMBH. Circles and solid lines indicate data from the first flare, with diamonds and dashed lines indicating data from second flares. We also include the density profile of Sw J1644+57 for reference. Data includes ASASSN-15oi \citep{hajela25-15oi}, AT2019ehz \citep{cendes24-mega} (this work), AT2019eve \citep{cendes24-mega} (this work), ASASSN-19bt \citep{christy24-19bt}, AT2019dsg \citep{cendes24-mega},  AT2020zso \citep{christy2025-20zso/21sdu}, AT2020vwl \citep{goodwin25-20vwl2},  and AT2024tvd \citep{sfaradi25-24tvd}. Sw J1644+57 \citep{berger12-j1644-1,zauderer13-j1644-2,eftekhari18-j1644-3,cendes21-j1644-4}.
\label{fig:density}}
\end{figure}

In Figure \ref{fig:density}, the circumnuclear medium density profiles are plotted against the radius, scaled by the Schwarzschild radius, $R_s$, with data for the first and second flares plotted as seen in Figure \ref{fig:energy-vel}.  The density profile traced by the first radio flares are often similar to those traced by the second flares, showing similar densities at similar distances (e.g. AT2020zso). This could be indicative of each outflow having varying geometries since it would be expected that a spherical blast wave would sweep the immediate medium as it moves, leaving a different profile density for the subsequent outflow to move through. AT2019ehz and AT2019eve have similar density profiles when compared to others in the sample, where most TDEs show densities $\propto R/R_s^{-1} -R/R_s^{-2} $. Specifically, AT2019ehz shows a profile on average $\propto R^{-2}$ while AT2019eve is more shallow at ${\sim} R^{-1}$. AT2019eve also traces out a profile to a much greater radius, which is likely due to its black hole mass being the estimated as the smallest in the sample at $10^{5.1\pm0.04} M_\odot$ \citep{vanVelzen21-tdediscs}, making the normalization of $R_s$ smaller in value than the rest of the sample. We note that this SMBH mass can vary substantially depending on the method used for estimation  (e.g., \verb|MOSFiT| estimates a black hole mass of $10^{6.4\pm0.2} M_\odot$; \citealt{alexander25-accretiondriven}), therefore the SMBH masses used here are each derived from properties of their host galaxies for consistency.

\section{Conclusions} \label{sect:conclusion}
We report the detection of two new secondary radio flares, or rebrightenings, in the TDEs AT2019ehz and AT2019eve. These are the only such rebrightenings from TDEs already exhibiting late-time flares.  

We summarize the radio behavior of AT2019ehz as follows:
\begin{itemize}
    \item Rebrightening was detected at 1545 days post-disruption, ${\sim}770$ days after the initial radio detection, with a luminosity thereafter of $2.7\times10^{38}$ erg s$^{-1}$.  This corresponds to a factor of $2\times$ increase from the previous observation of fading flare emission, which was $1.4\times10^{38}$ erg s$^{-1}$. 
    \item Energies of the outflow steadily increase from $2.9\times10^{47} -1.6\times10^{48}$ erg with velocities also increasing from $0.01c - 0.03c$.
    \item We favor a scenario of two separate outflows for this TDE, launched at roughly $400$ days and at $1070$ days post-disruption, due to delayed accretion activity or energy injection.
\end{itemize}

Additionally, we summarize the radio behavior of AT2019eve as follows:
\begin{itemize}
    \item  Rebrightening was detected at 1594 days post-disruption, ${\sim}820$ days after the initial radio detection, with luminosities ranging from $6.7\times10^{38}-1\times10^{39}$ erg sec$^{-1}$. This corresponds to a factor of $2-3\times$ increase from the previous observation of the fading flare emission at $3.6\times10^{38}$ erg s$^{-1}$. 
    \item The peak frequency $\nu_p$ increases from 1.6 - 2.8 GHz from 1325 - 2173 days, while $F_{\nu,p}$ is fairly consistent at roughly ${\sim}1.0$ mJy. 
    \item Outflow energy remains relatively constant at ${\sim}7\times10^{48}$ erg with velocities decreasing across epochs from $0.06c - 0.02c$ when using a prompt launch date.
    \item Due to the unusual evolution in $\nu_p$, no one scenario is favored as the origin of the second flare in this TDE, highlighting the need for more complicated models in TDE radio evolution.
\end{itemize} 

AT2019ehz and AT2019eve exhibit the latest radio rebrightenings observed to date when compared to other TDEs that have shown this behavior, with energy and density profiles are consistent with other rebrightening TDEs.  Both these sources are still evolving and continued monitoring is necessary to constrain the true origin of their rebrightenings.

It is clearly evident that TDEs continue to exhibit new and varied radio activity well into late-times, years after the initial disruption. Continued monitoring of all TDEs at these later times will help constrain the rates of multiple radio flares.  Additionally, late-time monitoring of TDEs is needed to understand the full rates for the diversity of TDE flares, and build a stronger understanding of TDE outflow evolution.

\begin{acknowledgments}
We thank Enrico Ramirez-Ruiz for helpful discussions on TDE outflows. We thank the National Radio Astronomy Observatory, a facility of the National Science Foundation operated under cooperative agreement by Associated Universities, Inc. NKV was supported by a SOS award from NRAO. AJG is grateful for support from the Forrest Research Foundation. 
\end{acknowledgments}

\bibliography{references}{}
\bibliographystyle{aasjournalv7}

\appendix

\section{Data}
\begin{longtable}{ccccccc}\toprule
Object&  Telescope&  Project Code&  Obs. Date&  $\delta$t (days)&  Freq. (GHz)& Flux Density (mJy)\\\midrule \endhead
AT2019ehz&  VLA&  23A-241&  2023-07-22&  1545&  1.26& 0.171 $\pm$ 0.014\\
 & ...& ...& ...& ...& 1.78&0.320 $\pm$ 0.019\\
 & ...& ...& ...& ...& 2.5&0.298 $\pm$ 0.009\\
 & ...& ...& ...& ...& 3.5&0.396 $\pm$ 0.007\\
 & ...& ...& ...& ...& 5.0&0.422 $\pm$ 0.013\\
 & ...& ...& ...& ...& 7.0&0.405 $\pm$ 0.010\\
 & ...& ...& ...& ...& 9.0&0.332 $\pm$ 0.008\\
 & ...& ...& ...& ...& 11.0&0.294 $\pm$ 0.012\\
         &  ...&  SC240292&  2023-09-12&  1597&  1.26& 0.139 $\pm$ 0.027\\
 & ...& ...& ...& ...& 1.78&0.228 $\pm$ 0.026\\
 & ...& ...& ...& ...& 2.5&0.341 $\pm$ 0.013\\
 & ...& ...& ...& ...& 3.5&0.415 $\pm$ 0.009\\
 & ...& ...& ...& ...& 5.0&0.332 $\pm$ 0.014\\
 & ...& ...& ...& ...& 7.0&0.296 $\pm$ 0.012\\
 & ...& ...& ...& ...& 9.0&0.302 $\pm$ 0.010\\
 & ...& ...& ...& ...& 10.0&0.266 $\pm$ 0.010\\
 & ...& ...& ...& ...& 11.0&$<0.035$\\
 & ...& ...& ...& ...& 13.55&0.205 $\pm$ 0.011\\
 & ...& ...& ...& ...& 16.62&0.077 $\pm$ 0.012\\
         &  ...&  24A-322&  2024-07-16&  1905&  1.26& 0.324 $\pm$ 0.015\\
 & ...& ...& ...& ...& 1.78&0.469 $\pm$ 0.018\\
 & ...& ...& ...& ...& 2.5&0.459 $\pm$ 0.012\\
 & ...& ...& ...& ...& 3.5&0.468 $\pm$ 0.009\\
 & ...& ...& ...& ...& 5.0&0.377 $\pm$ 0.015\\
 & ...& ...& ...& ...& 7.0&0.291 $\pm$ 0.010\\
 & ...& ...& ...& ...& 9.0&0.216 $\pm$ 0.010\\
 & ...& ...& ...& ...& 11.0&0.187 $\pm$ 0.013\\
         &  ...&  25A-159&  2025-07-09&  2263&  1.26& 0.462 $\pm$ 0.038\\
 & ...& ...& ...& ...& 1.78&0.534 $\pm$ 0.024\\
 & ...& ...& ...& ...& 2.5&0.360 $\pm$ 0.031\\
 & ...& ...& ...& ...& 3.5&0.423 $\pm$ 0.019\\
 & ...& ...& ...& ...& 5.0&0.393 $\pm$ 0.027\\
 & ...& ...& ...& ...& 7.0&0.261 $\pm$ 0.015\\
 & ...& ...& ...& ...& 9.0&0.273 $\pm$ 0.012\\
 & ...& ...& ...& ...& 11.0&0.187 $\pm$ 0.017\\
         &  ...&  26A-167&  2026-04-01&  2529&  1.26& 0.233 $\pm$ 0.028\\
 & ...& ...& ...& ...& 1.78&0.244 $\pm$ 0.022\\
 & ...& ...& ...& ...& 9.0&0.104 $\pm$ 0.009\\
 & ...& ...& ...& ...& 11.0&0.076 $\pm$ 0.014\\\bottomrule
 AT2019eve&  VLA&  23A-241&  2023-09-15&  1594&  1.26& 0.864 $\pm$ 0.029\\
 & ...& ...& ...& ...& 1.78&0.996 $\pm$ 0.029\\
 & ...& ...& ...& ...& 2.5&0.962 $\pm$ 0.022\\
 & ...& ...& ...& ...& 3.5&0.980 $\pm$ 0.014\\
 & ...& ...& ...& ...& 5.0&0.867 $\pm$ 0.010\\
 & ...& ...& ...& ...& 7.0&0.707 $\pm$ 0.010\\
 & ...& ...& ...& ...& 9.0&0.551 $\pm$ 0.013\\
 & ...& ...& ...& ...& 11.0&0.425 $\pm$ 0.009\\
         &  ...&  24A-322&  2024-08-23&  1937&  1.26& 0.717 $\pm$ 0.016\\
 & ...& ...& ...& ...& 1.78&0.855 $\pm$ 0.022\\
 & ...& ...& ...& ...& 2.5&1.130 $\pm$ 0.017\\
 & ...& ...& ...& ...& 3.5&1.125 $\pm$ 0.012\\
 & ...& ...& ...& ...& 5.0&0.926 $\pm$ 0.011\\
 & ...& ...& ...& ...& 7.0&0.702 $\pm$ 0.016\\
 & ...& ...& ...& ...& 9.0&0.533 $\pm$ 0.017\\
 & ...& ...& ...& ...& 11.0&0.221 $\pm$ 0.021\\
         &  ...&  25A-159&  2025-04-16&  2173&  1.26& $<0.741$\\
 & ...& ...& ...& ...& 1.78&1.155 $\pm$ 0.059\\
 & ...& ...& ...& ...& 2.5&0.947 $\pm$ 0.068\\
 & ...& ...& ...& ...& 3.5&1.053 $\pm$ 0.029\\
 & ...& ...& ...& ...& 5.0&1.022 $\pm$ 0.023\\
 & ...& ...& ...& ...& 7.0&0.943 $\pm$ 0.020\\
 & ...& ...& ...& ...& 9.0&0.698 $\pm$ 0.014\\
 & ...& ...& ...& ...& 11.0&0.648 $\pm$ 0.013\\
         &  ...&  25A-159&  2025-08-16&  2295&  5.0& 1.293 $\pm$ 0.018\\
 & ...& ...& ...& ...& 7.0&1.035 $\pm$ 0.015\\
         &  ...&  25B-154&  2025-10-03&  2343&  1.25& 1.346 $\pm$ 0.020\\
 & ...& ...& ...& ...& 1.75&1.458 $\pm$ 0.017\\
 & ...& ...& ...& ...& 9.0&0.740 $\pm$ 0.025\\
 & ...& ...& ...& ...& 11.0&0.589 $\pm$ 0.032\\\bottomrule
\caption{Radio flux density measurements}
\label{tab:fd-measurements}
    
\end{longtable}

\section{Interstellar Scintillation} \label{sect:scintillation}
We investigate the presence of interstellar scintillation as an alternative explanation of the flux variations seen across the light curves of AT2019ehz and AT2019eve. Using the NE2001 electron-density model \citep{cordes04-galelectrondensity,ocker24-pythonNE2001} at the Galactic coordinates of AT2019ehz, we find a transition frequency of 7.1 GHz between strong and weak scintillation regimes, a scattering measure of $SM=9.8\times10^{-5}$ kpc m$^{-20/3}$, and an effective screen distance of $d_{scr}\approx0.98$ kpc. At the Galactic coordinates of AT2019eve, we find a transition frequency of 7.0 GHz, $SM=1.2\times10^{-4}$ kpc m$^{-20/3}$, and $d_{scr}\approx0.71$ kpc. Using these parameters, we find the critical radius for diffractive scintillation to be $\theta_{crit}\approx3$ $\mu$as for both AT2019ehz and AT2019eve \citep{goodman97-scintillation}. We infer the angular radius to range between 8 - 24 $\mu$as for AT2019ehz and 21 - 36 $\mu$as for AT2019eve, suggesting diffractive scintillation would not be expected to affect measurements for either source.  

At these angular sizes, however flux measurements could be susceptible to variations due to refractive scintillation. Following \cite{walker98-scintillation}, we expect these sources will be in the strong, refractive scintillation regime until the source reaches an angular size of ${\sim}100$ $\mu$as. Modulation is expected to range from $\approx50-1\%$ across the observed frequency range, with the peak modulation at ${\sim}3$ GHz. 

We calculated the additional error with the appropriate modulation at each frequency by adding in quadrature to the flux density error and found that the variations seen at the rebrightenings across the frequencies are still larger than can be accounted for by these increased uncertainties. Therefore, we conclude these rebrightenings are not artifacts due to interstellar scintillation for either TDE.

\end{document}